\documentclass[superscriptaddress,showpacs,twocolumn,prd]{revtex4}
\usepackage{graphicx}

\begin{document}

%%%%%%%%%%%Extra added for preprint%%%%%%%%%%%%%%%%%%%%%%%%%%%%%%%%%%%%%%%%%
\onecolumngrid
\thispagestyle{empty}
\begin{flushright}
{\large 
LU TP 05-1\\
\large hep-lat/0501014\\[0.1cm]
\large January 2005}
\end{flushright}
\vskip5cm
\begin{center}
{\Large\bf
Decay Constants of Pseudoscalar Mesons to Two Loops \\[0.5cm]
       in Three-Flavor Partially Quenched $\chi$PT}

\vskip2cm

{\large \bf Johan Bijnens and Timo A. L\"ahde}\\[1cm]
{Department of Theoretical Physics, Lund University,\\
S\"olvegatan 14A, S 223 62 Lund, Sweden}

\vskip3cm

{\large\bf Abstract}

\vskip1cm

\parbox{14cm}{\large
This paper presents a first study of the decay
constants of the charged, or flavor-off-diagonal, pseudoscalar mesons
to two loops for three flavors of sea quarks, in Partially Quenched 
Chiral Perturbation Theory (PQ$\chi$PT). Explicit analytical expressions up 
to ${\cal O}(p^6)$ in the momentum expansion are given. The calculations have
been performed within the supersymmetric formulation of PQ$\chi$PT.
We also present some numerical results to indicate the size of the
corrections.}

\vskip2cm

\end{center}
\vskip2cm
\twocolumngrid
\setcounter{page}{0}
%%%%%%%%%%%%%%%%%%%%%%%%%%%%%%%%%%%%%%%%%%%%%%%%%%%%%%%%%%%%%%%%%%%%%%%%%%%%

\title{Decay Constants of Pseudoscalar Mesons to Two Loops \\
       in Three-Flavor Partially Quenched $\chi$PT
      }

\author{Johan Bijnens}
\affiliation{Department of Theoretical Physics, Lund University,\\
S\"olvegatan 14A, S 223 62 Lund, Sweden}
\author{Timo A. L\"ahde}
\affiliation{Department of Theoretical Physics, Lund University,\\
S\"olvegatan 14A, S 223 62 Lund, Sweden}


\begin{abstract} 
This paper presents a first study of the decay
constants of the charged, or flavor-off-diagonal, pseudoscalar mesons
to two loops for three flavors of sea quarks, in Partially Quenched 
Chiral Perturbation Theory (PQ$\chi$PT). Explicit analytical expressions up 
to ${\cal O}(p^6)$ in the momentum expansion are given. The calculations have
been performed within the supersymmetric formulation of PQ$\chi$PT.
We also present some numerical results to indicate the size of the
corrections.
\end{abstract}

\maketitle

\section{Introduction}

The derivation of low-energy hadronic observables, e.g. meson masses
and decay constants, from the theory of the strong interaction (QCD)
has so far proven to be impossible by means of analytical methods.
Because of this situation, numerical Lattice QCD simulations, whereby
the functional integral is evaluated numerically on a discretized
space-time lattice, have developed into a major field of study. Such
simulations are, however, seriously hampered by difficulties in the
simulation of dynamical sea quark effects.  Although much progress
has been made recently, it is still impractical, for computational
reasons, to simulate with sea quark masses that are close to the
physical $u,d$ quark masses of a few MeV. This situation, with sea
quark masses of a few tens of MeV, is referred to as partially
quenched~(PQ) QCD. Consequently, the physical values of the sea quark
masses have to be reached by extrapolation from the partially
quenched simulation results.

A practical method for this extrapolation is provided by Chiral
Perturbation Theory ($\chi$PT), which provides the correct quark mass
dependences of the various physical quantities that are measured on
the lattice. Standard three-flavor $\chi$PT as introduced by Weinberg,
Gasser and Leutwyler in Refs.~\cite{GL}, is valid in the
(unquenched) QCD case of equal valence and sea quark masses. The
generalization of $\chi$PT to the quenched case (without sea quarks) or
to the partially quenched case (sea quark masses different from the
valence ones) has been carried out by Bernard and Golterman in
Refs.~\cite{BG1,BG2}. The quark mass dependence of partially quenched
chiral perturbation theory (PQ$\chi$PT) is explicit, and thus the
limit where the sea quark masses become equal to the valence quark
masses can be taken. As a consequence, $\chi$PT is included in 
PQ$\chi$PT and the free parameters, or low-energy constants (LEC:s), 
of $\chi$PT can be directly obtained from those
of PQ$\chi$PT~\cite{BG2,Sharpe1}. 

The calculation of charged pseudoscalar meson masses and decay
constants to one loop (NLO) in PQ$\chi$PT has been carried out
in Refs.~\cite{BG2,Sharpe1,Sharpe2}, and first 
results for the mass of a charged pseudoscalar meson at two loops
or next-to-next-to-leading order (NNLO) in
PQ$\chi$PT have already appeared, for degenerate sea quark masses, in
Ref.~\cite{BDL}. The need for such calculations is clear as NNLO
effects have already been detected in Lattice QCD
simulations~\cite{Latt1,Latt2}. A calculation of the pseudoscalar
meson masses for nondegenerate sea quarks is in progress~\cite{BDL2}.

This paper presents the first calculation of the decay constants of the
charged, or flavor off-diagonal, pseudoscalar mesons in NNLO PQ$\chi$PT,
for three flavors of sea quarks ($n_{\mathrm{sea}} = 3$). The results 
are characterized by the number of nondegenerate valence
and sea quarks, denoted $d_{\mathrm{val}}$ and $d_{\mathrm{sea}}$,
respectively. For the decay constants of the charged pseudoscalar mesons, 
the maximum number of nondegenerate valence quark masses is 
$d_{\mathrm{val}} = 2$. The degree of quark mass degeneracy in each 
result is sometimes also referred to with the notation 
$d_{\mathrm{val}} + d_{\mathrm{sea}}$. The decay constant 
of the charged pion in the $SU(2)$ symmetric limit thus corresponds to the
$1 + 2$ case. Likewise, the decay constants of the
charged and neutral kaons may be obtained from the 
$d_{\mathrm{val}} = 2$ results with $d_{\mathrm{sea}}$ = 2. Results are
also presented for the case of $d_{\mathrm{sea}}$ = 1 (all sea 
quark masses equal), and $d_{\mathrm{sea}}$ = 3 (all sea quark masses 
different). An extension of the present work to the neutral 
pseudoscalar mesons is also planned.

The analytical expressions for the NNLO shift of the decay constants
are in general very long, but the expressions simplify considerably when pairs
of sea or valence quark masses become degenerate. In view of this, the NNLO 
loop results are given separately for each case of $d_{\mathrm{val}} +
d_{\mathrm{sea}}$ considered. In the next sections, the technical
background for the NNLO calculations, the full results for the decay
constants of the charged pseudoscalar mesons and numerical results as
a function of the input quark masses are given, along with a
concluding discussion.

\section{Technical Overview}

Most of the technical aspects that concern the calculation of the
pseudoscalar meson decay constants to two loops, or NNLO, are
identical to those of the calculation of the pseudoscalar meson mass,
and have already been justified in Ref.~\cite{BDL}. Most
significantly, the Lagrangians of PQ$\chi$PT at ${\cal O}(p^4)$ and
${\cal O}(p^6)$ may be directly obtained from the corresponding
Lagrangians of normal (unquenched) $n_F$ flavor $\chi$PT, provided
that the traces and meson matrices are replaced with the
supertraces and meson matrices relevant to the partially quenched
theory~\cite{BG1,BG2,Sharpe1}. This can be argued from the Replica method
as in Ref.~\cite{replica}, or by the fact that all the relations used
to constrain the number of terms in Ref.~\cite{BCE1} remain valid when
traces are changed to supertraces. We work here in the version of PQ$\chi$PT
without the $\Phi_0$ as discussed in Ref.~\cite{Sharpe2}.

\subsection{Masses and Low-Energy Constants}

All calculations in this paper have been performed with three flavors 
of valence quarks, three flavors of sea quarks and three flavors of 
bosonic 'ghost' quarks. These may be viewed as the $u,d$ and $s$
quarks in the valence, sea and ghost sectors, respectively. The purpose
of the ghost quarks is to remove the valence quark loops which are
disconnected from the external legs.

The input quark masses $m_i$ enter into the calculation in terms of the
lowest order squared meson masses $\chi_i$, which are defined as usual
in $\chi$PT, by $\chi_i = 2B_0 m_i$. In the present calculations, we thus
have three valence inputs $\chi_1,\chi_2,\chi_3$, three sea inputs
$\chi_4,\chi_5,\chi_6$, and three ghost inputs $\chi_7,\chi_8,\chi_9$. In
order for the disconnected valence quark loops to be canceled, the masses
of the ghost quarks are always equal to those of the corresponding valence
quarks, such that $\chi_7 = \chi_1$, $\chi_8 = \chi_2$ and 
$\chi_9 = \chi_3$. Explicitly, for $d_{\mathrm{val}}=1$, we have 
$\chi_1 = \chi_2 = \chi_3$, for $d_{\mathrm{val}}=2$, we have 
$\chi_1 = \chi_2 \ne \chi_3$ and for $d_{\mathrm{val}}=3$, we have 
$\chi_1 \ne \chi_2 \ne \chi_3$. Similarly, for the sea quarks
$d_{\mathrm{sea}}=1$ implies $\chi_4 = \chi_5 = \chi_6$, while
$d_{\mathrm{sea}}=2$ implies $\chi_4 = \chi_5 \ne \chi_6$
and finally $d_{\mathrm{sea}}=3$ that $\chi_4 \ne \chi_5 \ne \chi_6$.

The number of independent low-energy constants in unquenched and partially 
quenched $\chi$PT is slightly different, but the former are always linear 
combinations of the latter. For PQ$\chi$PT, they are $F_0$ and $B_0$ at 
leading order, $L^r_0$ through $L^r_{12}$ at NLO and $K^r_1$ through
$K^r_{115}$ at NNLO. In contrast, for three flavor unquenched $\chi$PT 
they are $F_0$ and $B_0$ at leading order, $L^{r(3)}_0$ through $L^{r(3)}_{12}$
at NLO and $C^r_1$ through $C^r_{94}$ at NNLO. Note that the parameters 
$L^r_{11,12}$ and $L^{r(3)}_{11,12}$ correspond to the usual $H_{1,2}^r$. Also,
the parameters $F_0$ and $B_0$ are identical for unquenched and partially
quenched $\chi$PT. At NLO, the relations between the low-energy constants
are~\cite{GL,BCE1,BCE2}
\begin{eqnarray}
L^{r(3)}_1 &=& L^r_0/2 + L^r_1, \nonumber \\
L^{r(3)}_2 &=& L^r_0 + L^r_2, \nonumber \\
L^{r(3)}_3 &=& -2 L^r_0 + L^r_3, \nonumber \\
L^{r(3)}_{4\ldots 12} &=& L^{r}_{4\ldots 12}\,,
\end{eqnarray}
and the corresponding linear relations relevant for the NNLO parameters
can be found in Ref.~\cite{BCE1}.

\subsection{Meson Propagators in PQ$\chi$PT}

The calculation of the pseudoscalar meson mass in Ref.~\cite{BDL} was 
only performed for $d_{\mathrm{val}} = d_{\mathrm{sea}}= 1$. However, the 
present calculation of the decay constants to NNLO is also concerned with the 
more general cases of $d_{\mathrm{val}}$ and $d_{\mathrm{sea}} = 2,3$. This 
leads to much more involved expressions because of the appearance of the 
residues of the neutral meson propagators. We recall here the results of 
Ref.~\cite{Sharpe2} which we have translated to Minkowski space from 
the Euclidean formalism used there.

In Minkowski space, the propagator $G_{ij}^c$ of a charged, or
flavor-off-diagonal meson with flavor structure $q_i\bar q_j$ in 
supersymmetric PQ$\chi$PT, is given by~\cite{Sharpe2}
\begin{eqnarray}
-i\,G_{ij}^c (k) &=& 
\frac{\epsilon_j}{k^2 - \chi_{ij} + i\varepsilon}\quad (i \neq j)\,.
\label{propc}
\end{eqnarray}
The factor $\chi_{ij}$ is defined in terms of the squared masses
as $\chi_{ij} = (\chi_i + \chi_j) / 2$. The sign vector $\epsilon_j$
is defined as $+1$ for the fermionic valence and sea quarks ($j=1,\ldots,6$)
and $-1$ for the bosonic ghost quarks ($j=7,8,9$).
The propagator of a flavor-neutral meson in supersymmetric
PQ$\chi$PT is more complicated, as it connects mesons of different flavor
indices as well. The propagator for a meson with quantum numbers
$q_i\bar q_i$ to one with quantum numbers $q_j\bar q_j$ may be written 
in Minkowski space~\cite{Sharpe2} as
\begin{eqnarray}
G_{ij}^n (k) &=& G_{ij}^c (k)\,\delta_{ij} 
- \frac{1}{n_{\mathrm{sea}}}\,G_{ij}^q (k).
\label{propn}
\end{eqnarray}

The nontrivial part $G_{ij}^q$ of the neutral meson propagator may
be expressed, by means of partial fractioning, in terms of a sum of
single and double poles in $k^2$. For the most general case of 
$d_{\mathrm{sea}} = 3$, in terms of the single-pole residue $R$, 
the double-pole residue $R^d$ and the auxiliary residue $R^c$, the 
propagator $G_{ij}^q$ is
\begin{eqnarray}
-i\,G_{ij}^q (k) &=& \frac{R^{i}_{j\pi\eta}}{k^2 - \chi_i + i\varepsilon}
+ \frac{R^{j}_{i\pi\eta}}{k^2 - \chi_j + i\varepsilon} \label{npropij} \\ 
&+& \frac{R_{\eta ij}^\pi} {k^2 - \chi_\pi + i\varepsilon} 
+ \frac{R_{\pi ij}^\eta}{k^2 - \chi_\eta + i\varepsilon} \quad (i \neq j),
\nonumber \\
-i\,G_{ii}^q (k) &=& \frac{R^d_i}{(k^2 - \chi_i + i\varepsilon)^2} 
+ \frac{R^c_i}{k^2 - \chi_i + i\varepsilon} \label{npropii} \\
&+& \frac{R_{\eta ii}^\pi} {k^2 - \chi_\pi + i\varepsilon} 
+ \frac{R_{\pi ii}^\eta}{k^2 - \chi_\eta + i\varepsilon}.
\nonumber
\end{eqnarray}
For uniformity of notation, the squared (lowest order) masses of
the neutral pion and the eta meson in the sea quark sector have been denoted by
$\chi_\pi$ and $\chi_\eta$, respectively. They are functions of the
sea quark masses, and in the case of $d_{\mathrm{sea}} = 3$ they are
given by the lowest order $\chi$PT result with $\pi^0 - \eta$ mixing
active~\cite{Sharpe2}.  They can be obtained from the relations
\begin{eqnarray}
\chi_\pi+\chi_\eta &=& \frac{2}{3}\left(\chi_4+\chi_5+\chi_6\right),
\nonumber\\
\chi_\pi \chi_\eta &=& \frac{1}{3}
\left(\chi_4\chi_5+\chi_5\chi_6+\chi_4\chi_6\right).
\end{eqnarray}

For $d_{\mathrm{sea}} = 2$, the $\pi^0$ pole
in $G_{ij}^q$ disappears, and in that case the (lowest order) eta meson
mass is trivially given in terms of the remaining sea quark masses as
$\chi_\eta = 1/3\,\chi_4 + 2/3\,\chi_6$. In this case the index $\pi$ 
has been suppressed in the residue notation. Further, for 
$d_{\mathrm{sea}} = 1$, both the $\pi^0$ and $\eta$ poles disappear and
consequently both the $\pi$ and $\eta$ indices have been suppressed. In this 
way we obtain a unique notation for the residues, from which the number of
nondegenerate sea quarks is immediately apparent.

The various residues $R$ of the propagators of the neutral, or flavor-diagonal 
mesons~\cite{Sharpe1,Sharpe2} appear in the results, and are one reason 
why the PQ$\chi$PT expressions are much more involved than the $\chi$PT 
results of Ref.~\cite{ABT1}. The use of the propagators~(\ref{npropij}) 
and~(\ref{npropii}) in the present form has the advantage of producing 
results in terms of standard loop integrals which can be treated with 
known methods. On the other hand, the various residues $R$ of the 
flavor-neutral meson propagator fulfill a very large
number of relations and the direct output from the calculations of the
diagrams consequently produces a large number of redundant
terms. This problem does not yet manifest itself at the one-loop
level, but becomes troublesome at the two-loop level when the mass
degeneracies in the sea and valence quark sectors are lifted.
With a major effort, the end result can be simplified, in
some cases it has been compressed by more than an order of magnitude.

\subsection{Notation for Propagator Residues}

The form of the single-pole residues $R_{jkl}^{i}$ and the
double-pole residue $R_i^d$, which appear in Eqs.~(\ref{npropij})
and~(\ref{npropii}), depends on the degree of degeneracy in the sea
quark masses, which in turn is indicated by the number of indices in the
single-pole residue $R$. It is useful to define the more general 
quantities $R^z_{a\ldots b}$ such that
\begin{eqnarray}
R^z_{ab} &=& \chi_a - \chi_b, \nonumber \\
R^z_{abc} &=& \frac{\chi_a - \chi_b}{\chi_a - \chi_c}, \nonumber \\ 
R^z_{abcd} &=& \frac{(\chi_a - \chi_b)(\chi_a - \chi_c)}
{\chi_a - \chi_d}, \nonumber \\
R^z_{abcdefg} &=& \frac{(\chi_a - \chi_b)(\chi_a - \chi_c)(\chi_a - \chi_d)}
{(\chi_a - \chi_e)(\chi_a - \chi_f)(\chi_a - \chi_g)},
\label{RSfunc}
\end{eqnarray}
and so on. The $R^z$ notation is primarily useful for defining the
residues of the flavor-neutral propagators, but it may also appear
independently in the final result for the decay constant. In such cases, the
$R^z$ have been generated by simplification procedures, as all the
residues that are naturally generated by partial fractioning of the propagator
$G_{ij}^q$ are of the form given below. 
Note that $R^z_{a\ldots b}$ has the same dimension as $\chi_i$ for
an even number of indices and is dimensionless for an odd number of indices.

For the case of $d_{\mathrm{sea}} = 1$, all residues associated with the sea
quark sector have reduced to numbers. Some nontrivial residues still appear if
the valence quarks are nondegenerate, according to
\begin{eqnarray}
R_{j}^{i} &=& R^z_{i4j}, \nonumber \\
R_{i}^{d} &=& R^z_{i4}\,.
\end{eqnarray}
For convenience, $R_{i}^{d}$ is also used for $d_{\mathrm{val}} = 1$ 
in order to maintain a notation similar to the more general cases.

As noted in Ref.~\cite{Sharpe2}, the residues also simplify for
$d_{\mathrm{sea}} = 2$. In particular, the residue of the 
neutral pion pole in the sea quark sector ($R^\pi_{jkl}$) vanishes, 
and the remaining ones satisfy a larger number of relations than for 
$d_{\mathrm{sea}} = 3$. The remaining nontrivial residues for $d_{\mathrm{sea}} = 2$
may be expressed as
\begin{eqnarray}
R_{jk}^{i} &=& R^z_{i46jk}, \nonumber \\
R_{i}^{d} &=& R^z_{i46\eta},\nonumber \\
R_{i}^{c} &=& R^i_{4\eta} + R^i_{6\eta} - R^i_{\eta\eta}.
\end{eqnarray}
It is also noteworthy that the above residue notation is highly redundant
because of the trivial relation between $\chi_4,\chi_6$ and $\chi_\eta$. 
This fact has been exploited in the simplification of the end results.

For $d_{\mathrm{sea}} = 3$, the 
naturally generated residues are
\begin{eqnarray}
R_{jkl}^{i} &=& R^z_{i456jkl}, \nonumber \\
R_{i}^{d} &=& R^z_{i456\pi\eta}, \nonumber \\
R_{i}^{c} &=& R^i_{4\pi\eta} + R^i_{5\pi\eta} + R^i_{6\pi\eta}
          - R^i_{\pi\eta\eta} - R^i_{\pi\pi\eta}.
\end{eqnarray}
Even in this case there are many relations between the various residues.
Some can be found in Ref.~\cite{Sharpe1} but many more exist. Finally, it
should be noted that in the limit where the sea quark masses become equal to
the valence quark masses, the propagator residues of PQ$\chi$PT reduce so that
the $\pi^0$ and $\eta$ meson propagators of unquenched $\chi$PT are 
recovered.

\subsection{Integral Notation}

The expression for the decay constant of a charged pseudoscalar meson
depends, in NNLO PQ$\chi$PT, on a number of one-loop and two-loop 
integrals. The finite parts of the chiral logarithms $A,B$ and $C$ are
\begin{eqnarray}
\bar A(\chi) &=& -\pi_{16}\, \chi \log(\chi/\mu^2), \nonumber \\
\bar B(\chi_i,\chi_j;0) &=& -\pi_{16}\, \frac{\chi_i\log(\chi_i/\mu^2) 
- \chi_j\log(\chi_j/\mu^2)}{\chi_i - \chi_j},
\nonumber\\
\bar C(\chi,\chi,\chi;0) &=& -\pi_{16}/(2 \chi)\, ,
\end{eqnarray}
where the subtraction scale dependence has been moved into the loop
integrals. We also define $\pi_{16} = 1/(16 \pi^2)$. 
Note that in the limit $\chi_i = \chi_j$, 
the integral $\bar B$ reduces to
\begin{eqnarray}
\bar B(\chi,\chi;0) &=& -\pi_{16}\left(1 + \log(\chi/\mu^2) \right).
\end{eqnarray}
The following combinations of finite one-loop integrals are also introduced,
as they
are naturally generated by the procedure of dimensional regularization:
\begin{eqnarray}
\bar A(\chi;\varepsilon) &=& \bar A(\chi)^2 / (2\pi_{16}\,\chi)
\nonumber \\
&+& \pi_{16}\,\chi\,(\pi^2/12 + 1/2), \nonumber \\
\bar B(\chi,\chi;0,\varepsilon) &=& 
\bar A(\chi)\bar B(\chi,\chi;0) / (\pi_{16}\,\chi)
\nonumber \\
&-& \bar A(\chi)^2 / (2\pi_{16}\,\chi^2) 
\nonumber \\
&+& \pi_{16}\,(\pi^2/12 + 1/2).
\end{eqnarray}
Additionally, the upper middle and upper
left diagrams in Fig.~\ref{decfig} generate integrals of the form
\begin{equation}
\bar B(\chi_i,\chi_j;0,k) = \chi_i \bar B(\chi_i,\chi_j;0) + \bar A(\chi_j),
\end{equation}
which are symmetric under the interchange of $\chi_i$ and $\chi_j$.
The above integrals have been introduced in order to make the
symmetries in the end results more explicit.

The finite two-loop integrals $H^F,H_1^F,H_{21}^F$ that are
generated by the top right diagram of Fig.~\ref{decfig} may be
evaluated using the methods of Ref.~\cite{ABT1}. Note that the
corresponding primed integrals $H^{F'},H_1^{F'},H_{21}^{F'}$ indicate
differentiation with respect to $p^2$. The notation used for the $H$
integrals in this paper is similar to that of Ref.~\cite{ABT1}, except that
an extra integer argument now indicates the propagator structure,
such that e.g.
\begin{equation}
H^F(\chi_i,\chi_j,\chi_k;p^2) \rightarrow H^F(n,\chi_i,\chi_j,\chi_k;p^2).
\end{equation}
The case of $n = 1$ indicates that the integral consists of
single propagators only, as in Ref.~\cite{ABT1}, whereas \mbox{$n =
2$} indicates that the first propagator appears squared and \mbox{$n =
3$} that the second propagator appears squared. The cases with two
double propagators that can appear in the calculations are \mbox{$n =
5$}, for which the first and second propagators appear squared, and
\mbox{$n = 7$} for which the second and third propagators are squared. 
Explicit expressions for \mbox{$n = 1$} can be found in
Ref.~\cite{ABT1}, and the other cases may be obtained by
differentiation with respect to the masses of those expressions.

\subsection{Simplification of End Results}

A number of combinations of quark masses and propagator residues are
naturally generated in the calculations, and have consequently been
given special notations. The most common of these is
\begin{eqnarray}
\bar\chi_n &=& \frac{1}{3}\sum_{i\,=\,4,5,6} \chi_i^n,
\end{eqnarray}
of which $\bar\chi_1$ is equal to the average sea quark mass $\bar\chi$
defined in Ref.~\cite{Sharpe1}. Other combinations that appear in the 
calculations consist of products of quark masses and propagator residues.
For $d_{\mathrm{sea}} = 2$, these include
\begin{eqnarray}
\bar\chi^{a}_{bn} &=& \frac{1}{3}\sum_{i\,=\,4,5,6} R^a_{bi}\,\chi_i^n, 
\nonumber \\
\bar\chi^{ac}_{bdn} &=& \frac{1}{3}\sum_{i\,=\,4,5,6}
    R^a_{bi}\,R^c_{di}\,\chi_i^n, 
\nonumber \\
\bar\chi^{ace}_{bdn} &=& \frac{1}{3}\sum_{i\,=\,4,5,6} R^a_{bi}\,R^c_{di}\,
R^e_{ii}\,\chi_i^n, 
\nonumber \\ 
R^v_{ijk} &=& R^i_{jj} + R^i_{kk} - 2 R^i_{jk}\,.
\end{eqnarray}
For $d_{\mathrm{sea}} = 3$ these become
\begin{eqnarray}
\bar\chi^{a}_{bn} &=& \frac{1}{3}\sum_{i\,=\,4,5,6} R^a_{bii}\,\chi_i^n, 
\nonumber \\
\bar\chi^{a}_{bcn} &=& \frac{1}{3}\sum_{i\,=\,4,5,6} R^a_{bci}\,\chi_i^n, 
\nonumber \\
\bar\chi^{ad}_{bcefn} &=& \frac{1}{3}\sum_{i\,=\,4,5,6}
 R^a_{bci}\,R^d_{efi}\,\chi_i^n, 
\nonumber \\
\bar\chi^{adk}_{bcefln} &=& \frac{1}{3}\sum_{i\,=\,4,5,6}
 R^a_{bci}\,R^d_{efi}\,
R^k_{lii}\,\chi_i^n, 
\nonumber \\
R^v_{ijkl} &=& R^i_{jkk} + R^i_{jll} - 2 R^i_{jkl}.
\end{eqnarray}
It should be noted that also in the case of $d_{\mathrm{sea}} = 2$,
the sums run over all three sea quark flavors. 
Furthermore, only some of the above $\bar\chi$ 
functions appear in the results, as many of them can be reexpressed
in terms of other $R$ or $R^z$ functions.

The actual simplification of the end results is at first glance a
formidable task, since the expression for $F_a$ as calculated from the
diagrams in Fig.~\ref{decfig} contains several thousand terms. 
Especially for $d_{\mathrm{val}} = 2$, the parts proportional to
$\bar A(\chi_\eta;\varepsilon)$ and $\bar A(\chi_1;\varepsilon)$
alone are several hundred terms long, while in simplified form they
typically contain only~$\sim 5$ terms. However, these simplifications
are not easily apparent and are consequently best accomplished by the
employment of suitable software, such as \verb;Maple; or 
\verb;Mathematica;. Even so, as the expressions can seldom be factored
into a single term, considerable trial and error is usually required.

\subsection{Summation Conventions}

The expressions for the decay constants are symmetric, within the sea 
and valence quark sectors, under the interchange of quark masses.
Consequently, the end results can be conveniently compactified by the
introduction of summation conventions which exploit these symmetries.
For example, the notation for the sea quark sector may be considerably 
compactified by the introduction of the summation indices $s$ and $t$, 
which may appear for the quark masses $\chi$, and among the indices 
of the functions $R$ and $\bar\chi$. 

These sea-quark summation indices should be interpreted as follows:
If an index $t$ is present once or several times, then there will 
always be an occurrence of the $s$ index as well,
and the term is then to be summed over all pairs of different sea 
quark indices. If the index $s$ is present but $t$ is not, then the 
entire term is to be summed over all sea quark indices. Simple examples 
of terms are
\begin{eqnarray}
\chi_s &=& \sum_{i\,=\,4,5,6} \chi_i, \nonumber \\
\chi_{st} &=& \sum_{{}^{i,j\,=\,4,5,6}_{j>i}} \chi_{ij}\,.
\end{eqnarray}
For terms consisting of products of several factors,
the summation sign should always be inserted at the
beginning of each relevant term, such that e.g.
\begin{eqnarray}
\bar A(\chi_s)\,R^1_{\eta s}\,\chi_s &=& \sum_{i\,=\,4,5,6}
\bar A(\chi_i)\,R^1_{\eta i}\,\chi_i.
\end{eqnarray}
Thus any contribution which is written in terms of the indices 
$s$ and $t$ explicitly fulfills the required symmetry properties in the 
sea quark sector. Note that also for $d_{\mathrm{sea}} = 
2$, the summation is over all three sea quark flavors, although the
index $t$ has not been implemented in that case. 

Further compactification of the results is possible, as the valence quark
sector is symmetric under the exchange of the valence quark indices 
$1$ and $3$. For this purpose, the summation indices $p$ and $q$ have 
been introduced. These are in the present paper
only needed for the case of $d_{\mathrm{val}} = 2$ and then always
occur for the squared valence quark masses $\chi_1$ and $\chi_3$.
If the index $q$ is present, there will always be an index $p$ and the
resulting sum is over the pairs $(p,q) = (1,3)$ and 
$(p,q) = (3,1)$. If only $p$ is present, the sum is over the indices 
$1$ and~$3$. An example is
\begin{eqnarray}
\bar A(\chi_p)\,R^p_{q\eta}\,\chi_p &=&
\bar A(\chi_1)\,R^1_{3\eta}\,\chi_1 \:+\: [1\leftrightarrow 3].
\end{eqnarray}
Any contribution written in terms of the $(p,q)$ notation is thus symmetric
under the 
interchange of the valence quark masses $\chi_1$ and $\chi_3$. 

For $d_{\mathrm{sea}} = 3$, there exists an additional symmetry, 
i.e. the results are symmetric under the interchange of the lowest
order neutral meson masses, $\chi_\pi$ and $\chi_\eta$, in the sea-quark 
sector. This is exploited by the summation indices $m$ and $n$. If the 
index $m$ is present, there will always be an index $n$ and the 
corresponding sum is over the pairs $(m,n) = (\pi,\eta)$ and 
$(m,n) = (\eta,\pi)$. An example is
\begin{eqnarray}
\bar A(\chi_m)\,R^m_{n11}\,\chi_m &=&
\bar A(\chi_\eta)\,R^{\eta}_{\pi 11}\,\chi_\eta \:+\: 
[\eta\leftrightarrow \pi].
\end{eqnarray}
The above summation conventions thus provide a means of eliminating
a large number of terms in the expression for $F_a$, which are of 
similar form. As an added bonus, it gives a possibility 
to conveniently check that the NNLO loop results fulfill the
required symmetry relations in the sea and valence sectors.

\section{The Pseudoscalar Meson Decay Constants to ${\cal O}(p^6)$}

\begin{figure}[h!]
\begin{center}
\includegraphics[width=\columnwidth]{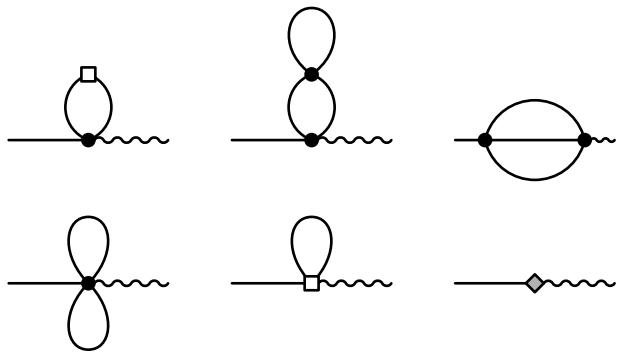}
\caption{Feynman diagrams at ${\mathcal O}(p^6)$ or two-loop for the
matrix element of the axial current operator. Filled circles denote
vertices of the ${\mathcal L}_2$ Lagrangian, whereas open squares and
shaded diamonds denote vertices of the ${\mathcal L}_4$ and
${\mathcal L}_6$ Lagrangians, respectively.}
\label{decfig}
\end{center}
\end{figure}

The decay constants $F_a$ of the pseudoscalar mesons are obtained
from the definition
\begin{equation}
\langle 0| A_a^\mu(0) |\phi_a(p)\rangle = i\sqrt{2}\,p^\mu\,F_a,
\label{decdef}
\end{equation}
in terms of the axial current operator $A_a^\mu$. The diagrams that
contribute to that operator at ${\mathcal O}(p^6)$, or NNLO, are shown in
Fig.~\ref{decfig}. Diagrams at ${\mathcal O}(p^2)$ and ${\mathcal
O}(p^4)$ also contribute to Eq.~(\ref{decdef}) via the
renormalization of the pseudoscalar meson wave function $\phi_a(p)$. The
results
for charged pseudoscalar mesons so obtained depend on the 
${\cal O}(p^6)$ low-energy constants $K_{19}^r$ through $K_{23}^r$. 

The results are expressed in terms of the valence inputs $\chi_1,\chi_3$ 
and the sea inputs $\chi_4,\chi_5,\chi_6$, which are defined in terms of
the quark masses through $\chi_i = 2 B_0 m_i$, and the quantity $\chi_{ij} =
(\chi_i+\chi_j)/2$, which corresponds to the lowest order charged
meson mass. Other parameters include the decay constant in the chiral
limit ($F_0$), the quark condensate in the chiral limit, via
$\langle{\bar q}q\rangle = - B_0 F_0^2$, and the LEC:s of ${\cal
O}(p^4)$ and ${\cal O}(p^6)$, i.e. the $L_i^r$ and the 
$K_i^r$~\cite{BCE1,BCE2}, respectively.

The decay constants of the pseudoscalar mesons are given in the form
\begin{equation}
F_{\mathrm{phys}} = F_0 \left[ 1 + \frac{\delta^{(4)\mathrm{vs}}}{F_0^2} 
+ \frac{\delta^{(6)\mathrm{vs}}_{\mathrm{ct}} 
+ \delta^{(6)\mathrm{vs}}_{\mathrm{loops}}}{F_0^4}
+ \mathcal{O}(p^8) \right],
\label{delteq}
\end{equation}
where the ${\cal O}(p^4)$ and ${\cal O}(p^6)$ contributions have been
separated. The NNLO contribution $\delta^{(6)}$ has been further
split into the contributions from the chiral loops and from the
${\cal O}(p^6)$ counterterms. The superscripts (v) and (s) indicate the
values of $d_{\mathrm{val}}$ and $d_{\mathrm{sea}}$, respectively.

\subsection{Results for $d_{\mathrm{val}} = 1$}

The NLO result for $d_{\mathrm{val}} = 1$ is fairly short, and will thus
only be given for $d_{\mathrm{sea}} = 3$. The results for 
$d_{\mathrm{sea}} = 1,2$ may readily be derived from that expression. The 
combined NLO result (loops and counterterms) for $d_{\mathrm{sea}} = 3$, 
which is in agreement with Ref.~\cite{Sharpe2}, is
\begin{equation} 
\delta^{(4)13} = 12\,L^r_{4}\,\bar\chi_{1}
\:+\: 4\,L^r_{5}\,\chi_1 
\:+\: 1/2\,\bar{A}(\chi_{1s}).
\label{F0p413} 
\end{equation} 

At NNLO, the chiral loops form, by far, the largest
contribution to the decay constant. As a straightforward derivation
of the results for $d_{\mathrm{sea}} = 1,2$ from the
$d_{\mathrm{sea}} = 3$ case is tedious and complicated, the expressions for 
the different cases are given separately below. As expected, the infinities in
all expressions for the decay constant have canceled. The appearance
of 'unphysical' $\bar B$ logarithms in the results is, in part, due to the 
partial quenching, and to the fact that the results have been expressed in 
terms of the lowest order masses rather 
than the full physical ones. Consequently, not all of them correspond
to the quenched chiral logarithms which are 
ill-behaved in the chiral limit. 

The contribution from the 
${\cal O}(p^6)$ counterterms to the decay constant at NNLO is, 
for $d_{\mathrm{sea}} = 3$,
\begin{eqnarray} 
\delta^{(6)13}_{\mathrm{ct}} & = & 8\,K_{19}^r\,\chi_1^2
\:+\: 24\,K_{20}^r\,\bar\chi_{1} \chi_1 
\:+\: 24\,K_{21}^r\,\bar\chi_{2} \nonumber \\
&+& 72\,K_{22}^r\,\bar\chi_{1}^2
\:+\: 8\,K_{23}^r\,\chi_1^2, 
\label{F0p613tree} 
\end{eqnarray} 
from which the results for $d_{\mathrm{sea}} = 1,2$ may be readily
inferred. The NNLO contributions from the chiral loops for 
$d_{\mathrm{sea}} = 1,2,3$ are, respectively,
\begin{widetext}
\begin{eqnarray} 
\delta^{(6)11}_{\mathrm{loops}} & = & 
\pi_{16}\,L^r_{0}\,
\left[ - 13/3\,\chi_1 \chi_4 + 1/2\,\chi_1^2 - 3/2\,\chi_4^2 \right]
\:-\,2\,\pi_{16}\,L^r_{1}\,\chi_1^2
\:-\: \pi_{16}\,L^r_{2}\,\left[ \chi_1^2 + 8\,\chi_4^2 \right] \nonumber \\
&+& \pi_{16}\,L^r_{3}\,\left[ - 17/6\,\chi_1 \chi_4 + 5/4\,\chi_1^2 - 3/4\,\chi_4^2 \right]
\:+\, \pi_{16}^2\,\left[ - 1/2\,\chi_1 \chi_4 + 1/64\,\chi_1^2 - 73/128\,\chi_4^2 \right] 
\nonumber \\ 
&-& 48\,L^r_{4}L^r_{5}\,\chi_1 \chi_4
\:-\,72\,L^{r2}_{4}\,\chi_4^2
\:-\,8\,L^{r2}_{5}\,\chi_1^2
\:+\,4\,\bar{A}(\chi_1)\,L^r_{0}\,\left[ \chi_1 + R^d_{1} \right]
\:-\,4\,\bar{A}(\chi_1)\,L^r_{1}\,\chi_1
\:-\,10\,\bar{A}(\chi_1)\,L^r_{2}\,\chi_1 
\nonumber \\
&+& 4\,\bar{A}(\chi_1)\,L^r_{3}\,\left[ \chi_1 + R^d_{1} \right]
\:-\,4/3\,\bar{A}(\chi_1)\,L^r_{5}\,\chi_1
\:-\,1/2\,\bar{A}(\chi_1) \bar{B}(\chi_{14},\chi_{14};0)\,\chi_{14}
\:+\,1/8\,\bar{A}(\chi_1;\varepsilon)\,\pi_{16}\,\chi_4 
\nonumber \\
&-& \bar{A}(\chi_{14})\,\pi_{16}\,\left[ 3/4\,\chi_{14} + 9/8\,\chi_4 \right]
\:-\,12\,\bar{A}(\chi_{14})\,L^r_{0}\,\chi_{14}
\:-\,30\,\bar{A}(\chi_{14})\,L^r_{3}\,\chi_{14}
\:-\,18\,\bar{A}(\chi_{14})\,L^r_{4}\,\chi_4 
\nonumber \\
&+& 6\,\bar{A}(\chi_{14})\,L^r_{5}\,\chi_1
\:+\,9/4\,\bar{A}(\chi_{14};\varepsilon)\,\pi_{16}\,\chi_4
\:-\,64\,\bar{A}(\chi_4)\,L^r_{1}\,\chi_4
\:-\,16\,\bar{A}(\chi_4)\,L^r_{2}\,\chi_4
\:+\,32\,\bar{A}(\chi_4) L^r_{4}\,\chi_4 
\nonumber \\
&+& \bar{A}(\chi_4;\varepsilon)\,\pi_{16}\,\chi_4
\:+\,4\,\bar{B}(\chi_1,\chi_1;0)\,L^r_{0}\,R^d_{1}\,\chi_1
\:+\,4\,\bar{B}(\chi_1,\chi_1;0)\,L^r_{3}\,R^d_{1}\,\chi_1
\:-\,4/3\,\bar{B}(\chi_1,\chi_1;0)\,L^r_{5}\,R^d_{1}\,\chi_1 \nonumber \\
&-& 1/8\,\bar{B}(\chi_1,\chi_1;0,\varepsilon)\,\pi_{16}\,R^d_{1}\,\chi_1
\:-\,36\,\bar{B}(\chi_{14},\chi_{14};0)\,L^r_{4}\,\chi_{14} \chi_4
\:-\,12\,\bar{B}(\chi_{14},\chi_{14};0)\,L^r_{5}\,\chi_{14}^2 \nonumber \\
&+& 72\,\bar{B}(\chi_{14},\chi_{14};0)\,L^r_{6}\,\chi_{14} \chi_4
\:+\,24\,\bar{B}(\chi_{14},\chi_{14};0)\,L^r_{8}\,\chi_{14}^2
\:-\,1/8\,H^{F}(1,\chi_1,\chi_{14},\chi_{14};\chi_1)\,\left[ \chi_1 - R^d_{1} \right]
\nonumber \\
&-& H^{F}(1,\chi_{14},\chi_{14},\chi_4;\chi_1)\,\chi_4
\:+\,1/8\,H^{F}(2,\chi_1,\chi_{14},\chi_{14};\chi_1)\,R^d_{1}\,\chi_1
\:+\,5/18\,H^{F'}(1,\chi_1,\chi_1,\chi_1;\chi_1)\,\chi_1^2 \nonumber \\
&+& H^{F'}(1,\chi_1,\chi_{14},\chi_{14};\chi_1)\,[ 1/8\,\chi_1 \chi_4 - 1/2\,\chi_1^2 ] 
\:+\, H^{F'}(1,\chi_{14},\chi_{14},\chi_4;\chi_1)\,\chi_1 \chi_4 \nonumber \\
&+& 2/9\,H^{F'}(2,\chi_1,\chi_1,\chi_1;\chi_1)\,R^d_{1}\,\chi_1^2
\:+\,3/8\,H^{F'}(2,\chi_1,\chi_{14},\chi_{14};\chi_1)\,R^d_{1}\,\chi_1^2
\:+\,1/9\,H^{F'}(5,\chi_1,\chi_1,\chi_1;\chi_1)\,(R^{d}_{1})^2 \chi_1^2 \nonumber \\
&-& 2\,H^{F'}_1(3,\chi_{14},\chi_1,\chi_{14};\chi_1)\,R^d_{1} \chi_1^2 
\:+\,3/8\,H^{F'}_{21}(1,\chi_1,\chi_{14},\chi_{14};\chi_1)\,\chi_1^2 
\:+\,3\,H^{F'}_{21}(1,\chi_4,\chi_{14},\chi_{14};\chi_1)\,\chi_1^2 \nonumber \\
&-& 3/8\,H^{F'}_{21}(2,\chi_1,\chi_{14},\chi_{14};\chi_1)\,R^d_{1}\,\chi_1^2,
\label{F0p611loop} \\
&& \nonumber \\
\delta^{(6)12}_{\mathrm{loops}} & = & 
\pi_{16}\,L^r_{0}\,\left[ 4/9\,\chi_\eta \chi_4 + 1/2\,\chi_1^2 
 - 13/3\,\bar\chi_{1} \chi_1 - 35/18\,\bar\chi_{2} \right]
\:-\,2\,\pi_{16}\,L^r_{1}\,\chi_1^2 \nonumber \\
&-& \pi_{16}\,L^r_{2}\,\left[ 11/3\,\chi_\eta \chi_4 + \chi_1^2 
+ 13/3\,\bar\chi_{2} \right]
\:+\,\pi_{16}\,L^r_{3}\,\left[ 4/9\,\chi_\eta \chi_4 + 5/4\,\chi_1^2 
- 17/6\,\bar\chi_{1} \chi_1 - 43/36\,\bar\chi_{2} \right] \nonumber \\ 
&+& \pi_{16}^2\,\left[ - 15/64\,\chi_\eta \chi_4 + 1/64\,\chi_1^2 
- 1/2\,\bar\chi_{1} \chi_1 - 43/128\,\bar\chi_{2} \right]
\:-\,48\,L^r_{4}L^r_{5}\,\bar\chi_{1}\,\chi_1 
\:-\,72\,L^{r2}_{4}\,\bar\chi_{1}^2 
\:-\,8\,L^{r2}_{5}\,\chi_1^2 \nonumber \\
&+& 4\,\bar{A}(\chi_\eta)\,L^r_{0}\,R^{\eta}_{11}\,\chi_\eta 
\:-\,8\,\bar{A}(\chi_\eta)\,L^r_{1}\,\chi_\eta
\:-\,2\,\bar{A}(\chi_\eta)\,L^r_{2}\,\chi_\eta
\:+\,4\,\bar{A}(\chi_\eta)\,L^r_{3}\,R^{\eta}_{11}\,\chi_\eta
\:+\,4\,\bar{A}(\chi_\eta)\,L^r_{4}\,\chi_\eta \nonumber \\
&-& 4/3\,\bar{A}(\chi_\eta)\,L^r_{5}\,R^{\eta}_{11}\,\chi_1
\:-\,1/6\,\bar{A}(\chi_\eta) \bar{B}(\chi_{1s},\chi_{1s};0)\,R^{\eta}_{1s}\,\chi_{1s}
\:+\,1/8\,\bar{A}(\chi_\eta;\varepsilon)\,\pi_{16}\,R^c_{1}\,\chi_\eta \nonumber \\
&+& 4\,\bar{A}(\chi_1)\,L^r_{0}\,\left[ R^c_{1}\,\chi_1 + R^d_{1} \right]
\:-\,4\,\bar{A}(\chi_1)\,L^r_{1}\,\chi_1
\:-\,10\,\bar{A}(\chi_1)\,L^r_{2}\,\chi_1
\:+\, 4\,\bar{A}(\chi_1)\,L^r_{3}\,\left[ R^c_{1}\,\chi_1 + R^d_{1} \right] \nonumber \\
&-& 4/3\,\bar{A}(\chi_1)\,L^r_{5}\,R^c_{1}\,\chi_1
\:-\,1/6\,\bar{A}(\chi_1) \bar{B}(\chi_{1s},\chi_{1s};0)\,R^{1}_{s\eta}\,\chi_{1s}  
\:+\,\bar{A}(\chi_1;\varepsilon)\,\pi_{16}\,\left[ 1/4\,\chi_1 - 1/8\,R^c_{1} \chi_1 
- 1/8\,R^d_{1} \right] \nonumber \\
&-& 24\,\bar{A}(\chi_4)\,L^r_{1}\,\chi_4
\:-\,6\,\bar{A}(\chi_4)\,L^r_{2}\,\chi_4
\:+\,12\,\bar{A}(\chi_4)\,L^r_{4}\,\chi_4
\:+\,3/8\,\bar{A}(\chi_4;\varepsilon)\,\pi_{16}\,\chi_4
\:-\,32\,\bar{A}(\chi_{46})\,L^r_{1}\,\chi_{46} \nonumber \\
&-& 8\,\bar{A}(\chi_{46})\,L^r_{2}\,\chi_{46}
\:+\,16\,\bar{A}(\chi_{46})\,L^r_{4}\,\chi_{46}
\:+\,1/2\,\bar{A}(\chi_{46};\varepsilon)\,\pi_{16}\,\chi_{46}
\:-\,\bar{A}(\chi_{1s})\,\pi_{16}\,\left[ 1/4\,\chi_{1s} + 3/8\,\bar\chi_{1} \right] 
\nonumber \\
&-& 4\,\bar{A}(\chi_{1s})\,L^r_{0}\,\chi_{1s}
\:-\,10\,\bar{A}(\chi_{1s})\,L^r_{3}\,\chi_{1s}
\:-\,6\,\bar{A}(\chi_{1s})\,L^r_{4}\,\bar\chi_{1}
\:+\,2\,\bar{A}(\chi_{1s})\,L^r_{5}\,\chi_1
\:+\,3/8\,\bar{A}(\chi_{1s};\varepsilon)\,\pi_{16}\,\left[\chi_s + \bar\chi_{1} \right]
\nonumber \\ 
&+& 4\,\bar{B}(\chi_1,\chi_1;0)\,L^r_{0}\,R^d_{1}\,\chi_1
\:+\,4\,\bar{B}(\chi_1,\chi_1;0)\,L^r_{3}\,R^d_{1}\,\chi_1
\:-\,4/3\,\bar{B}(\chi_1,\chi_1;0)\,L^r_{5}\,R^d_{1}\,\chi_1 \nonumber \\
&-& 1/8\,\bar{B}(\chi_1,\chi_1;0,\varepsilon)\,\pi_{16}\,R^d_{1}\,\chi_1
\:-\,12\,\bar{B}(\chi_{1s},\chi_{1s};0)\,L^r_{4}\,\bar\chi_{1}\,\chi_{1s}
\:-\,4\,\bar{B}(\chi_{1s},\chi_{1s};0)\,L^r_{5}\,\chi_{1s}^2 \nonumber \\
&+& 24\,\bar{B}(\chi_{1s},\chi_{1s};0)\,L^r_{6}\,\bar\chi_{1}\,\chi_{1s}
\:+\,8\,\bar{B}(\chi_{1s},\chi_{1s};0)\,L^r_{8}\,\chi_{1s}^2
\:+\,1/24\,H^{F}(1,\chi_\eta,\chi_{1s},\chi_{1s};\chi_1)\,R^v_{\eta 1s}\,\chi_\eta \nonumber \\ 
&+& H^{F}(1,\chi_1,\chi_{1s},\chi_{1s};\chi_1)\,\left[ - 1/12\,R^{1}_{s\eta}\,\chi_1 
+ 1/24\,R^c_{1}\,\chi_1 + 1/24\,R^d_{1} \right]
\:-\,3/8\,H^{F}(1,\chi_{14},\chi_{14},\chi_4;\chi_1)\,\chi_4 \nonumber \\
&-& 1/2\,H^{F}(1,\chi_{14},\chi_{16},\chi_{46};\chi_1)\,\chi_{46}
\:+\,1/24\,H^{F}(2,\chi_1,\chi_{1s},\chi_{1s};\chi_1)\,R^d_{1}\,\chi_1 \nonumber \\
&+& 1/9\,H^{F'}(1,\chi_\eta,\chi_\eta,\chi_1;\chi_1)\,(R^{\eta}_{11})^2 \chi_1^2 
\:+\,2/9\,H^{F'}(1,\chi_\eta,\chi_1,\chi_1;\chi_1)\,R^{\eta}_{11} R^c_{1}\,\chi_1^2 
\nonumber \\
&-& H^{F'}(1,\chi_\eta,\chi_{1s},\chi_{1s};\chi_1)\,\left[ 1/6\,R^{\eta}_{11}\,\chi_1^2 
+ 1/24\,R^v_{\eta 1s}\,\chi_\eta \chi_1 \right] 
\:+\,H^{F'}(1,\chi_1,\chi_1,\chi_1;\chi_1)\,\left[ 1/6\,\chi_1^2 \right. \nonumber \\
&+& 1/9 \left. (R^{c}_{1})^2 \chi_1^2 \right] 
\:+\,H^{F'}(1,\chi_1,\chi_{1s},\chi_{1s};\chi_1)\,\left[ - 1/4\,R^{1}_{s\eta}\,\chi_1^2 
+ 1/8\,R^c_{1}\,\chi_1^2 - 1/24\,R^d_{1}\,\chi_1 \right] \nonumber \\
&+& 3/8\,H^{F'}(1,\chi_{14},\chi_{14},\chi_4;\chi_1)\,\chi_1 \chi_4
\:+\,1/2\,H^{F'}(1,\chi_{14},\chi_{16},\chi_{46};\chi_1)\,\chi_1 \chi_{46} \nonumber \\
&+& 2/9\,H^{F'}(2,\chi_1,\chi_\eta,\chi_1;\chi_1)\,R^{\eta}_{11} R^d_{1}\,\chi_1^2 
\:+\,2/9\,H^{F'}(2,\chi_1,\chi_1,\chi_1;\chi_1)\,R^c_{1} R^d_{1}\,\chi_1^2 \nonumber \\ 
&+& 1/8\,H^{F'}(2,\chi_1,\chi_{1s},\chi_{1s};\chi_1)\,R^d_{1}\,\chi_1^2
\:+\,1/9\,H^{F'}(5,\chi_1,\chi_1,\chi_1;\chi_1)\,(R^{d}_{1})^2\,\chi_1^2 \nonumber \\
&-& 1/3\,H^{F'}_1(1,\chi_\eta,\chi_{1s},\chi_{1s};\chi_1)\,R^{\eta}_{1s} R^z_{1s\eta}\,
\chi_1^2
\:-\,2/3\,H^{F'}_1(1,\chi_{1s},\chi_{1s},\chi_1;\chi_1)\,R^{\eta}_{1s} R^z_{1s\eta}\,
\chi_1^2 \nonumber \\
&-& 2/3\,H^{F'}_1(3,\chi_{1s},\chi_1,\chi_{1s};\chi_1)\,R^d_{1}\,\chi_1^2
\:-\,1/8\,H^{F'}_{21}(1,\chi_\eta,\chi_{1s},\chi_{1s};\chi_1)\,R^v_{\eta 1s}\,\chi_1^2 
\nonumber \\
&+& H^{F'}_{21}(1,\chi_1,\chi_{1s},\chi_{1s};\chi_1)\,\left[ 1/4\,R^{1}_{s\eta}\,\chi_1^2 
- 1/8\,R^c_{1}\,\chi_1^2 \right]
\:+\,9/8\,H^{F'}_{21}(1,\chi_4,\chi_{14},\chi_{14};\chi_1)\,\chi_1^2 \nonumber \\
&+& 3/2\,H^{F'}_{21}(1,\chi_{46},\chi_{14},\chi_{16};\chi_1)\,\chi_1^2
\:-\,1/8\,H^{F'}_{21}(2,\chi_1,\chi_{1s},\chi_{1s};\chi_1)\,R^d_{1}\,\chi_1^2, 
\label{F0p612loop} \\
&& \nonumber \\
\delta^{(6)13}_{\mathrm{loops}} & = &
  \pi_{16}\,L^r_{0}\,\left[ 4/9\,\chi_\pi \chi_\eta + 1/2\,\chi_1^2 - 13/3\,\bar\chi_{1} 
\chi_1 - 35/18\,\bar\chi_{2} \right]
\:-\,2\,\pi_{16}\,L^r_{1}\,\chi_1^2 \nonumber \\
&-& \pi_{16}\,L^r_{2}\,\left[ 11/3\,\chi_\pi \chi_\eta + \chi_1^2 
+ 13/3\,\bar\chi_{2} \right]
\:+\,\pi_{16}\,L^r_{3}\,\left[ 4/9\,\chi_\pi \chi_\eta + 5/4\,\chi_1^2 - 17/6\,\bar\chi_{1} 
\chi_1 - 43/36\,\bar\chi_{2} \right] \nonumber \\
&+& \pi_{16}^2\,\left[ - 15/64\,\chi_\pi \chi_\eta + 1/64\,\chi_1^2 - 1/2\,\bar\chi_{1} 
\chi_1 - 43/128\,\bar\chi_{2} \right]
\:-\,48\,L^r_{4}L^r_{5}\,\bar\chi_{1} \chi_1
\:-\,72\,L^{r2}_{4}\,\bar\chi_{1}^2
\:-\,8\,L^{r2}_{5}\,\chi_1^2 \nonumber \\
&+& 4\,\bar{A}(\chi_m)\,L^r_{0}\,R^{m}_{n11}\,\chi_m
\:+\,8\,\bar{A}(\chi_m)\,L^r_{1}\,\bar\chi^{m}_{n0}\,\chi_m
\:+\,2\,\bar{A}(\chi_m)\,L^r_{2}\,\bar\chi^{m}_{n0}\,\chi_m
\:+\,4\,\bar{A}(\chi_m)\,L^r_{3}\,R^{m}_{n11}\,\chi_m \nonumber \\
&-& 4\,\bar{A}(\chi_m)\,L^r_{4}\,\bar\chi^{m}_{n1}
\:-\,4/3\,\bar{A}(\chi_m)\,L^r_{5}\,R^{m}_{n11}\,\chi_1
\:-\,1/6\,\bar{A}(\chi_m) \bar{B}(\chi_{1s},\chi_{1s},0)\,R^{m}_{n1s}\,\chi_{1s} \nonumber \\ 
&+& \bar{A}(\chi_m;\varepsilon)\,\pi_{16}\,\left[ - 7/24\,\bar\chi^{m}_{n0}\,\chi_m 
+ 1/6\,\bar\chi^{m}_{n1} - 1/8\,R^{m}_{n11}\,\chi_m \right]
\:+\,4\,\bar{A}(\chi_1)\,L^r_{0}\,\left[ R^c_{1}\,\chi_1 + R^d_{1} \right] \nonumber \\ 
&-& 4\,\bar{A}(\chi_1)\,L^r_{1}\,\chi_1
\:-\,10\,\bar{A}(\chi_1)\,L^r_{2}\,\chi_1
\:+\,4\,\bar{A}(\chi_1)\,L^r_{3}\,\left[ R^c_{1}\,\chi_1 + R^d_{1} \right]
\:-\,4/3\,\bar{A}(\chi_1)\,L^r_{5}\,R^c_{1}\,\chi_1 \nonumber \\
&-& 1/6\,\bar{A}(\chi_1) \bar{B}(\chi_{1s},\chi_{1s},0)\,R^{1}_{s\pi\eta}\,\chi_{1s}
\:+\,\bar{A}(\chi_1;\varepsilon)\,\pi_{16}\,\left[ 1/4\,\chi_1 - 1/8\,R^c_{1}\,\chi_1 
- 1/8\,R^d_{1} \right] \nonumber \\
&-& \bar{A}(\chi_{1s})\,\pi_{16}\,\left[ 1/4\,\chi_{1s} + 3/8\,\bar\chi_{1} \right] 
\:-\,4\,\bar{A}(\chi_{1s})\,L^r_{0}\,\chi_{1s}
\:-\,10\,\bar{A}(\chi_{1s})\,L^r_{3}\,\chi_{1s}
\:-\,6\,\bar{A}(\chi_{1s})\,L^r_{4}\,\bar\chi_{1} \nonumber \\
&+& 2\,\bar{A}(\chi_{1s})\,L^r_{5}\,\chi_1
\:+\,3/8\,\bar{A}(\chi_{1s};\varepsilon)\,\pi_{16}\,\left[ \chi_s + \bar\chi_{1} \right] 
\:+\,\bar{A}(\chi_s)\,L^r_{1}\,\left[ - 8\,\chi_s + 8/3\,R^c_{s}\,\chi_s \right] \nonumber \\
&-& \bar{A}(\chi_s)\,L^r_{2}\,\left[ 2\,\chi_s - 2/3\,R^c_{s}\,\chi_s \right]
\:+\,\bar{A}(\chi_s)\,L^r_{4}\,\left[ 4\,\chi_s - 4/3\,R^c_{s}\,\chi_s \right]
\:+\,\bar{A}(\chi_s;\varepsilon)\,\pi_{16}\,\left[ 1/8\,\chi_s - 1/24\,R^c_{s}\,\chi_s \right] 
\nonumber \\
&-& 16\,\bar{A}(\chi_{st})\,L^r_{1}\,\chi_{st}
\:-\,4\,\bar{A}(\chi_{st})\,L^r_{2}\,\chi_{st}
\:+\,8\,\bar{A}(\chi_{st})\,L^r_{4}\,\chi_{st}
\:+\,1/4\,\bar{A}(\chi_{st};\varepsilon)\,\pi_{16}\,\chi_{st} \nonumber \\
&+& 4\,\bar{B}(\chi_1,\chi_1;0)\,L^r_{0}\,R^d_{1}\,\chi_1
\:+\,4\,\bar{B}(\chi_1,\chi_1;0)\,L^r_{3}\,R^d_{1}\,\chi_1
\:-\,4/3\,\bar{B}(\chi_1,\chi_1;0)\,L^r_{5}\,R^d_{1}\,\chi_1 \nonumber \\ 
&-& 1/8\,\bar{B}(\chi_1,\chi_1;0,\varepsilon)\,\pi_{16}\,R^d_{1}\,\chi_1 
\:-\,12\,\bar{B}(\chi_{1s},\chi_{1s};0)\,L^r_{4}\,\bar\chi_{1}\,\chi_{1s}
\:-\,4\,\bar{B}(\chi_{1s},\chi_{1s};0)\,L^r_{5}\,\chi_{1s}^2 \nonumber \\
&+& 24\,\bar{B}(\chi_{1s},\chi_{1s};0)\,L^r_{6}\,\bar\chi_{1}\,\chi_{1s} 
\:+\,8\,\bar{B}(\chi_{1s},\chi_{1s};0)\,L^r_{8}\,\chi_{1s}^2
\:+\,1/24\,H^{F}(1,\chi_m,\chi_{1s},\chi_{1s};\chi_1)\,R^v_{mn1s}\,\chi_m \nonumber \\
&+& H^{F}(1,\chi_1,\chi_{1s},\chi_{1s};\chi_1)\,\left[ - 1/12\,R^{1}_{s\pi\eta}\,\chi_1 
+ 1/24\,R^c_{1}\,\chi_1 + 1/24\,R^d_{1} \right]
\:+\,H^{F}(1,\chi_{1s},\chi_{1s},\chi_s;\chi_1)\,\left[ - 1/8\,\chi_s \right. \nonumber \\
&+& 1/24 \left. R^c_{s}\,\chi_s \right]
\:-\,1/4\,H^{F}(1,\chi_{1s},\chi_{1t},\chi_{st};\chi_1)\,\chi_{st}
\:+\,1/24\,H^{F}(2,\chi_1,\chi_{1s},\chi_{1s};\chi_1)\,R^d_{1}\,\chi_1 \nonumber \\
&+& 1/9\,H^{F'}(1,\chi_m,\chi_m,\chi_1;\chi_1)\,(R^{m}_{n11})^2 \chi_1^2 
\:+\,2/9\,H^{F'}(1,\chi_m,\chi_1,\chi_1;\chi_1)\,R^{m}_{n11} R^c_{1}\,\chi_1^2 \nonumber \\ 
&-& H^{F'}(1,\chi_m,\chi_{1s},\chi_{1s};\chi_1)\,\left[ 1/6\,R^{m}_{n11}\,\chi_1^2 
+ 1/24\,R^v_{mn1s}\,\chi_m \chi_1 \right] \nonumber \\
&+& 2/9\,H^{F'}(1,\chi_\pi,\chi_\eta,\chi_1;\chi_1)\,R^{\pi}_{\eta 11} R^{\eta}_{\pi 11} 
\,\chi_1^2 
\:+\,H^{F'}(1,\chi_1,\chi_1,\chi_1;\chi_1)\,\left[ 1/6\,\chi_1^2 + 
1/9\,(R^{c}_{1})^2 \chi_1^2 \right] \nonumber \\
&+& H^{F'}(1,\chi_1,\chi_{1s},\chi_{1s};\chi_1)\,\left[ - 1/4\,R^{1}_{s\pi\eta}\,\chi_1^2 
+ 1/8\,R^c_{1}\,\chi_1^2 - 1/24\,R^d_{1}\,\chi_1 \right]
\:+\, H^{F'}(1,\chi_{1s},\chi_{1s},\chi_s;\chi_1)\,\left[ 1/8\,\chi_1 \chi_s \right. 
\nonumber \\ 
&-& 1/24 \left. R^c_{s}\,\chi_1 \chi_s \right] 
\:+\,1/4\,H^{F'}(1,\chi_{1s},\chi_{1t},\chi_{st};\chi_1)\,\chi_1 \chi_{st}
\:+\,2/9\,H^{F'}(2,\chi_1,\chi_m,\chi_1;\chi_1)\,R^{m}_{n11} R^d_{1}\,\chi_1^2 \nonumber \\ 
&+& 2/9\,H^{F'}(2,\chi_1,\chi_1,\chi_1;\chi_1)\,R^c_{1} R^d_{1}\,\chi_1^2 
\:+\,1/8\,H^{F'}(2,\chi_1,\chi_{1s},\chi_{1s};\chi_1)\,R^d_{1}\,\chi_1^2 \nonumber \\
&+& 1/9\,H^{F'}(5,\chi_1,\chi_1,\chi_1;\chi_1)\,(R^{d}_{1})^2\,\chi_1^2
\:-\,1/3\,H^{F'}_1(1,\chi_m,\chi_{1s},\chi_{1s};\chi_1)\,R^{m}_{n1s}\,R^z_{1sm}\,\chi_1^2
\nonumber \\
&-& 2/3\,H^{F'}_1(1,\chi_{1s},\chi_{1s},\chi_1;\chi_1)\,R^{m}_{n1s} R^z_{1sm}\,\chi_1^2 
\:-\,2/3\,H^{F'}_1(3,\chi_{1s},\chi_1,\chi_{1s};\chi_1)\,R^d_{1}\,\chi_1^2 \nonumber \\ 
&-& 1/8\,H^{F'}_{21}(1,\chi_m,\chi_{1s},\chi_{1s};\chi_1)\,R^v_{mn1s}\,\chi_1^2
\:+\,H^{F'}_{21}(1,\chi_1,\chi_{1s};\chi_{1s};\chi_1)\,\left[ 1/4\,R^{1}_{s\pi\eta}\,\chi_1^2 
- 1/8\,R^c_{1}\,\chi_1^2 \right] 
\nonumber \\
&+& H^{F'}_{21}(1,\chi_s,\chi_{1s},\chi_{1s};\chi_1)\,\left[ 3/8\,\chi_1^2 - 1/8\,R^c_{s} 
\,\chi_1^2 \right]
\:+\,3/4\,H^{F'}_{21}(1,\chi_{st},\chi_{1s},\chi_{1t};\chi_1)\,\chi_1^2 
\nonumber \\
&-& 1/8\,H^{F'}_{21}(2,\chi_1,\chi_{1s},\chi_{1s};\chi_1)\,R^d_{1}\,\chi_1^2. 
\label{F0p613loop} 
\end{eqnarray} 
\end{widetext}

\subsection{Results for $d_{\mathrm{val}} = 2$}

In general, the expressions for $d_{\mathrm{val}} = 2$ are
longer than those for
$d_{\mathrm{val}} = 1$, as the number of independent integrals that
can appear is 
significantly larger. The full result for $d_{\mathrm{sea}} = 3$ at
NNLO is very large,
and has not yet been worked out. Consequently, the largest number
of nondegenerate sea
quarks considered for $d_{\mathrm{val}} = 2$ is $d_{\mathrm{sea}} = 2$.
The NLO result for $d_{\mathrm{val}} = 2$ and $d_{\mathrm{sea}} = 2$, which
agrees with the result of 
Ref.~\cite{Sharpe2} is
\begin{eqnarray} 
\delta^{(4)22} & = &
 12\,L^r_{4}\,\bar\chi_{1}
\:+\,4\,L^r_{5}\,\chi_{13}
\:+\,\bar{A}(\chi_p)\,\left[ 1/6\,R^{p}_{q\eta} \right. \nonumber \\
&-& 1/12 \left. R^c_{p} \right] 
\:+\,1/4\,\bar{A}(\chi_{ps})
\:-\,1/12\,\bar{A}(\chi_\eta)\,R^v_{\eta 13} \nonumber \\
&-& 1/12\,\bar{B}(\chi_p,\chi_p,0)\,R^d_{p}.
\label{F0p422} 
\end{eqnarray} 

At NNLO, the contribution to the decay constant
from the ${\cal O}(p^6)$ counterterms for 
$d_{\mathrm{val}} = 2$ is similar to that for $d_{\mathrm{val}} = 1$. 
For $d_{\mathrm{sea}} = 
2$, that contribution is
\begin{eqnarray} 
\delta^{(6)22}_{\mathrm{ct}} & = & 
  4\,K^r_{19}\,\chi_p^2 
\:+\,24\,K^r_{20}\,\bar\chi_{1} \chi_{13}
\:+\,24\,K^r_{21}\,\bar\chi_{2} \nonumber \\
&+& 72\,K^r_{22}\,\bar\chi_{1}^2
\:+\,8\,K^r_{23}\,\chi_1 \chi_3.
\label{F0p622tree} 
\end{eqnarray} 

The NNLO contributions to the decay constant from the chiral loops are,
for $d_{\mathrm{sea}} 
= 1$ and $d_{\mathrm{sea}} = 2$, respectively,
\begin{widetext}

\begin{eqnarray} 
\delta^{(6)21}_{\mathrm{loops}} & = & 
 \pi_{16}\,L^r_{0}\,\left[ - 1/2\,\chi_1 \chi_3 - 13/3\,\chi_{13} \chi_4 + \chi_{13}^2 
- 3/2\,\chi_4^2 \right] 
\:-\,2\,\pi_{16}\,L^r_{1}\,\chi_{13}^2
\:-\,\pi_{16}\,L^r_{2}\,\left[ \chi_{13}^2 + 8\,\chi_4^2 \right] \nonumber \\
&+& \pi_{16}\,L^r_{3}\,\left[ - 7/12\,\chi_1 \chi_3 - 17/6\,\chi_{13} \chi_4 
+ 11/6\,\chi_{13}^2 - 3/4\,\chi_4^2 \right]
\:+\,\pi_{16}^2\,\left[ - 59/384\,\chi_1 \chi_3 - 1/2\,\chi_{13} \chi_4 + 65/384\,\chi_{13}^2 
\right. \nonumber \\
&-& 73/128 \left. \chi_4^2 \right]
\:-\,48\,L^r_{4}L^r_{5}\,\chi_{13} \chi_4
\:-\,72\,L^{r2}_{4}\,\chi_4^2
\:-\,8\,L^{r2}_{5}\,\chi_{13}^2
\:+\,\bar{A}(\chi_p)\,\pi_{16}\,\left[ 5/96\,\chi_p + 1/32\,\chi_q + 1/48\,\chi_4 \right.
\nonumber \\ 
&-& 1/12 \left. R^{p}_{q}\,\chi_p - 1/8\,R^{p}_{q}\,\chi_4 \right]
\:+\,\bar{A}(\chi_p)\,L^r_{0}\,\left[ 2/3\,\chi_p + 8/3\,R^{p}_{q}\,\chi_p 
+ 2/3\,R^d_{p} \right]
\:+\,\bar{A}(\chi_p)\,L^r_{3}\,\left[ 5/3\,\chi_p + 2/3\,R^{p}_{q}\,\chi_p \right. \nonumber \\
&+& 5/3 \left. R^d_{p} \right] 
\:+\,\bar{A}(\chi_p)\,L^r_{4}\,\left[ \chi_4 - 2\,R^{p}_{q}\,\chi_4 \right]
\:+\,\bar{A}(\chi_p)\,L^r_{5}\,\left[ 1/6\,\chi_p - 1/6\,\chi_q - 1/3\,\chi_4 - 2/3\,R^{p}_{q} 
\,\chi_p \right] \nonumber \\
&+& \bar{A}(\chi_p)^2\,\left[ 19/288 - 1/72\,R^{p}_{q} R^{q}_{p} \right]
\:-\,\bar{A}(\chi_p) \bar{A}(\chi_{p4})\,\left[ 1/16 + 1/12\,R^{p}_{q} \right]
\:-\,\bar{A}(\chi_p) \bar{A}(\chi_{q4})\,\left[ 3/16 + 1/12\,R^{p}_{q} \right] \nonumber \\
&+& 1/8\,\bar{A}(\chi_p) \bar{A}(\chi_{13})
\:+\,\bar{A}(\chi_p) \bar{B}(\chi_p,\chi_p;0)\,\left[ 11/36\,\chi_p - 1/18\,R^{p}_{q}\,\chi_p 
- 1/72\,R^{p}_{q} R^d_{p} + 1/144\,R^d_{p} \right] \nonumber \\
&+& \bar{A}(\chi_p) \bar{B}(\chi_q,\chi_q;0)\,\left[ - 1/72\,R^{p}_{q} R^d_{q} + 
1/144\,R^d_{q} \right]
\:-\,1/4\,\bar{A}(\chi_p) \bar{B}(\chi_{p4},\chi_{p4};0)\,\chi_{p4} \nonumber \\
&-& 1/18\,\bar{A}(\chi_p) \bar{B}(\chi_1,\chi_3;0)\,R^{q}_{p}\,\chi_p 
\:+\,1/18\,\bar{A}(\chi_p) \bar{C}(\chi_p,\chi_p,\chi_p;0)\,R^d_{p}\,\chi_p
\:+\,1/8\,\bar{A}(\chi_p;\varepsilon)\,\pi_{16}\,\left[ - \chi_p + R^{p}_{q}\,\chi_4 \right] 
\nonumber \\ 
&+& \bar{A}(\chi_{p4})\,\pi_{16}\,\left[ 3/16\,\chi_{p4} - 9/16\,\chi_{q4} 
- 9/16\,\chi_4 \right] 
\:-\,6\,\bar{A}(\chi_{p4})\,L^r_{0}\,\chi_{p4}
\:-\,15\,\bar{A}(\chi_{p4})\,L^r_{3}\,\chi_{p4}
\:-\,9\,\bar{A}(\chi_{p4})\,L^r_{4}\,\chi_4 \nonumber \\
&+& 3\,\bar{A}(\chi_{p4})\,L^r_{5}\,\chi_{13} 
\:-\,9/32\,\bar{A}(\chi_{p4})^2
\:+\,\bar{A}(\chi_{p4}) \bar{B}(\chi_p,\chi_p;0)\,\left[ 1/8\,\chi_p - 5/8\,\chi_{p4} \right] 
\:-\,1/16\,\bar{A}(\chi_{p4}) \bar{B}(\chi_q,\chi_q;0)\,R^d_{q} \nonumber \\
&+& 1/6\,\bar{A}(\chi_{p4}) \bar{B}(\chi_1,\chi_3;0)\,\chi_4 
\:+\,1/3\,\bar{A}(\chi_{p4}) \bar{B}(\chi_1,\chi_3;0,k)
\:+\,9/8\,\bar{A}(\chi_{p4};\varepsilon)\,\pi_{16}\,\chi_4
\:+\,\bar{A}(\chi_1) \bar{A}(\chi_3)\,\left[ - 1/144 \right. \nonumber \\
&+& 1/36 \left. R^{1}_{3} R^{3}_{1} \right]
\:-\,4\,\bar{A}(\chi_{13})\,L^r_{1}\,\chi_{13} 
\:-\,10\,\bar{A}(\chi_{13})\,L^r_{2}\,\chi_{13} 
\:+\,1/8\,\bar{A}(\chi_{13})^2 
\:-\,1/2\,\bar{A}(\chi_{13}) \bar{B}(\chi_1,\chi_3;0,k) \nonumber \\
&+& 1/4\,\bar{A}(\chi_{13};\varepsilon)\,\pi_{16}\,\chi_{13}
\:+\,9/16\,\bar{A}(\chi_{14}) \bar{A}(\chi_{34})
\:-\,64\,\bar{A}(\chi_4)\,L^r_{1}\,\chi_4
\:-\,16\,\bar{A}(\chi_4)\,L^r_{2}\,\chi_4
\:+\,32\,\bar{A}(\chi_4)\,L^r_{4}\,\chi_4 \nonumber \\
&+& 2/9\,\bar{A}(\chi_4) \bar{B}(\chi_p,\chi_p;0)\,\chi_4 
\:-\,4/9\,\bar{A}(\chi_4) \bar{B}(\chi_1,\chi_3;0)\,\chi_4 
\:+\,\bar{A}(\chi_4;\varepsilon)\,\pi_{16}\,\chi_4
\:+\,\bar{B}(\chi_p,\chi_p;0)\,\pi_{16}\,\left[ 1/96\,R^d_{p}\,\chi_p \right. \nonumber \\
&+& 1/32 \left. R^d_{p}\,\chi_q + 1/16\,R^d_{p}\,\chi_4 \right]
\:+\,2/3\,\bar{B}(\chi_p,\chi_p;0)\,L^r_{0}\,R^d_{p}\,\chi_p
\:+\,5/3\,\bar{B}(\chi_p,\chi_p;0)\,L^r_{3}\,R^d_{p}\,\chi_p \nonumber \\
&+& \bar{B}(\chi_p,\chi_p;0)\,L^r_{4}\,\left[ 2\,\chi_p \chi_4 - 4\,R^{p}_{q}\,\chi_p \chi_4 
+ 3\,R^d_{p}\,\chi_4 \right]
\:+\,\bar{B}(\chi_p,\chi_p;0)\,L^r_{5}\,\left[ - 2/3\,\chi_p \chi_4 + 4/3\,\chi_p^2 
- 4/3\,R^{p}_{q}\,\chi_p \chi_{13} \right. \nonumber \\
&-& 1/3 \left. R^d_{p}\,\chi_{13} \right]
\:+\,\bar{B}(\chi_p,\chi_p;0)\,L^r_{6}\,\left[ 4\,\chi_4^2 - 8\,R^{q}_{p}\,\chi_p \chi_4 \right]
\:+\,4\,\bar{B}(\chi_p,\chi_p;0)\,L^r_{7}\,(R^{d}_{p})^2
\:+\,\bar{B}(\chi_p,\chi_p;0)\,L^r_{8}\,\left[ 4/3\,\chi_4^2 \right. \nonumber \\
&-& 8/3 \left. R^{q}_{p}\,\chi_p^2 \right] 
\:+\,\bar{B}(\chi_p,\chi_p;0)^2\,\left[ 1/18\,R^{q}_{p} R^d_{p}\,\chi_p 
+ 1/288\,(R^{d}_{p})^2 \right] 
\:-\,1/18\,\bar{B}(\chi_p,\chi_p;0) \bar{B}(\chi_1,\chi_3;0)\,R^{q}_{p} R^d_{p}\,\chi_p 
\nonumber \\ 
&+& 1/18\,\bar{B}(\chi_p,\chi_p;0) \bar{C}(\chi_p,\chi_p,\chi_p;0)\,(R^{d}_{p})^2\,\chi_p 
\:-\,1/8\,\bar{B}(\chi_p,\chi_p;0,\varepsilon)\,\pi_{16}\,R^d_{p}\,\chi_{p4} \nonumber \\
&-& 18\,\bar{B}(\chi_{p4},\chi_{p4};0)\,L^r_{4}\,\chi_{p4} \chi_4 
\:-\,6\,\bar{B}(\chi_{p4},\chi_{p4};0)\,L^r_{5}\,\chi_{p4}^2
\:+\,36\,\bar{B}(\chi_{p4},\chi_{p4};0)\,L^r_{6}\,\chi_{p4} \chi_4 \nonumber \\
&+& 12\,\bar{B}(\chi_{p4},\chi_{p4};0)\,L^r_{8}\,\chi_{p4}^2
\:+\,1/144\,\bar{B}(\chi_1,\chi_1;0) \bar{B}(\chi_3,\chi_3;0)\,R^d_{1} R^d_{3} 
\:-\,8\,\bar{B}(\chi_1,\chi_3;0)\,L^r_{7}\,R^d_{1} R^d_{3} \nonumber \\ 
&-& 8/3\,\bar{B}(\chi_1,\chi_3;0)\,L^r_{8}\,R^d_{1} R^d_{3}
\:+\,4\,\bar{C}(\chi_p,\chi_p,\chi_p;0)\,L^r_{4}\,R^d_{p}\,\chi_p \chi_4
\:+\,4/3\,\bar{C}(\chi_p,\chi_p,\chi_p;0)\,L^r_{5}\,R^d_{p}\,\chi_p^2 \nonumber \\
&-& 8\,\bar{C}(\chi_p,\chi_p,\chi_p;0)\,L^r_{6}\,R^d_{p}\,\chi_p \chi_4
\:-\,8/3\,\bar{C}(\chi_p,\chi_p,\chi_p;0)\,L^r_{8}\,R^d_{p}\,\chi_p^2 
\:+\,H^{F}(1,\chi_p,\chi_p,\chi_{13};\chi_{13})\,\left[1/8\,\chi_p \right. \nonumber \\
&-& 5/72 \left. \chi_{13} + 1/36\,R^{p}_{q} R^{q}_{p}\,\chi_{13} \right]
\:+\,H^{F}(1,\chi_p,\chi_{14},\chi_{34};\chi_{13})\,\left[ - 1/16\,\chi_p + 1/16\,\chi_q 
- 1/8\,R^{p}_{q}\,\chi_4 \right] \nonumber \\
&+& H^{F}(1,\chi_1,\chi_{13},\chi_3;\chi_{13})\,\left[ 1/72\,\chi_{13} - 1/18\,R^{1}_{3} 
R^{3}_{1}\,\chi_{13} \right]
\:-\,1/8\,H^{F}(1,\chi_{13},\chi_{13},\chi_{13};\chi_{13})\,\chi_{13} \nonumber \\
&-& H^{F}(1,\chi_{14},\chi_{34},\chi_4;\chi_{13})\,\chi_4
\:+\,H^{F}(2,\chi_p,\chi_p,\chi_{13};\chi_{13})\,\left[ 1/36\,R^{p}_{q} R^d_{p}\,\chi_{13} 
- 1/72\,R^d_{p}\,\chi_{13} \right] \nonumber \\
&+& H^{F}(2,\chi_p,\chi_{13},\chi_q;\chi_{13})\,\left[ 1/36\,R^{q}_{p} R^d_{p}\,\chi_{13} 
- 1/72\,R^d_{p}\,\chi_{13} \right]
\:+\,1/8\,H^{F}(2,\chi_p,\chi_{14},\chi_{34};\chi_{13})\,R^d_{p}\,\chi_{p4} \nonumber \\
&-& 1/144\,H^{F}(5,\chi_p,\chi_p,\chi_{13};\chi_{13})\,(R^{d}_{p})^2\,\chi_{13} 
\:-\,1/72\,H^{F}(5,\chi_1,\chi_3,\chi_{13};\chi_{13})\,R^d_{1} R^d_{3}\,\chi_{13} \nonumber \\
&+& H^{F'}(1,\chi_p,\chi_p,\chi_{13};\chi_{13})\,\left[ - 1/8\,\chi_p \chi_{13} 
- 5/24\,\chi_{13}^2 + 1/12\,R^{p}_{q}\,\chi_{13}^2 - 1/36\,(R^{p}_{q})^2\,\chi_{13}^2 
\right] \nonumber \\
&+& H^{F'}(1,\chi_p,\chi_{14},\chi_{34};\chi_{13})\,\left[ 3/16\,\chi_p \chi_{13} 
- 1/16\,\chi_{13} \chi_4 - 1/8\,R^{p}_{q} R^d_{p}\,\chi_{13} + 3/8\,R^{q}_{p}\,\chi_{13}^2 
\right] \nonumber \\ 
&+& H^{F'}(1,\chi_1,\chi_{13},\chi_3;\chi_{13})\,\left[ - 1/72\,\chi_{13}^2 
+ 1/6\,R^{1}_{3} R^{3}_{1}\,\chi_{13}^2 \right]
\:+\,1/8\,H^{F'}(1,\chi_{13},\chi_{13},\chi_{13};\chi_{13})\,\chi_{13}^2 \nonumber \\ 
&+& H^{F'}(1,\chi_{14},\chi_{34},\chi_4;\chi_{13})\,\chi_{13} \chi_4
\:+\,H^{F'}(2,\chi_p,\chi_p,\chi_{13};\chi_{13})\,\left[ - 1/36\,R^{p}_{q} 
R^d_{p}\,\chi_{13}^2 + 5/72\,R^d_{p}\,\chi_{13}^2 \right] \nonumber \\
&+& H^{F'}(2,\chi_p,\chi_{13},\chi_q;\chi_{13})\,\left[ - 1/36\,R^{q}_{p} R^d_{p}\,\chi_{13}^2 
+ 1/72\,R^d_{p}\,\chi_{13}^2 \right]
\:-\,1/8\,H^{F'}(2,\chi_p,\chi_{14},\chi_{34};\chi_{13})\,R^d_{p}\,\chi_{p4} \chi_{13} 
\nonumber \\
&+& 5/144\,H^{F'}(5,\chi_p,\chi_p,\chi_{13};\chi_{13})\,(R^{d}_{p})^2\,\chi_{13}^2
\:+\,1/72\,H^{F'}(5,\chi_1,\chi_3,\chi_{13};\chi_{13})\,R^d_{1} R^d_{3}\,\chi_{13}^2 
\nonumber \\
&+& H^{F'}_1(1,\chi_p,\chi_p,\chi_{13};\chi_{13})\,\left[ 10/9\,\chi_{13}^2 - 2/9\,R^{p}_{q} 
R^{q}_{p}\,\chi_{13}^2 \right]
\:-\,H^{F'}_1(1,\chi_p,\chi_{14},\chi_{34};\chi_{13})\,R^{q}_{p}\,\chi_{13}^2 \nonumber \\
&-& H^{F'}_1(1,\chi_{p4},\chi_{q4},\chi_p;\chi_{13})\,\chi_{13}^2
\:+\,H^{F'}_1(1,\chi_{13},\chi_1,\chi_3;\chi_{13})\,\left[ 1/9\,\chi_{13}^2 - 2/9\,R^{1}_{3} 
R^{3}_{1}\,\chi_{13}^2 \right] \nonumber \\
&+& H^{F'}_1(3,\chi_{13},\chi_p,\chi_p;\chi_{13})\,\left[ 1/9\,R^{p}_{q} R^d_{p}\,\chi_{13}^2 
- 1/9\,R^d_{p}\,\chi_{13}^2 \right]
\:+\,1/9\,H^{F'}_1(3,\chi_{13},\chi_p,\chi_q;\chi_{13})\,R^{q}_{p} R^d_{p}\,\chi_{13}^2 
\nonumber \\
&-& 1/18\,H^{F'}_1(7,\chi_{13},\chi_p,\chi_p;\chi_{13})\,(R^{d}_{p})^2\,\chi_{13}^2
\:-\,3/8\,H^{F'}_{21}(1,\chi_p,\chi_p,\chi_{13};\chi_{13})\,\chi_{13}^2 \nonumber \\
&+& 3/8\,H^{F'}_{21}(1,\chi_p,\chi_{14},\chi_{34};\chi_{13})\,R^{q}_{p}\,\chi_{13}^2
\:+\,3/8\,H^{F'}_{21}(1,\chi_{p4},\chi_p,\chi_{q4};\chi_{13})\,R^{p}_{q}\,\chi_{13}^2
\nonumber \\
&-& 3/8\,H^{F'}_{21}(1,\chi_{p4},\chi_q,\chi_{q4};\chi_{13})\,R^{p}_{q}\,\chi_{13}^2  
\:+\,H^{F'}_{21}(1,\chi_{13},\chi_p,\chi_p;\chi_{13})\,\left[ 5/24\,\chi_{13}^2 
- 1/12\,R^{p}_{q} R^{q}_{p}\,\chi_{13}^2 \right] \nonumber \\
&+& H^{F'}_{21}(1,\chi_{13},\chi_1,\chi_3;\chi_{13})\,\left[ - 1/24\,\chi_{13}^2 
+ 1/6\,R^{1}_{3} R^{3}_{1}\,\chi_{13}^2 \right]
\:+\,3/8\,H^{F'}_{21}(1,\chi_{13},\chi_{13},\chi_{13};\chi_{13})\,\chi_{13}^2 \nonumber \\
&+& 3\,H^{F'}_{21}(1,\chi_4,\chi_{14},\chi_{34};\chi_{13})\,\chi_{13}^2
\:-\,3/8\,H^{F'}_{21}(3,\chi_{p4},\chi_p,\chi_{q4};\chi_{13})\,R^d_{p}\,\chi_{13}^2 \nonumber \\
&+& H^{F'}_{21}(3,\chi_{13},\chi_p,\chi_p;\chi_{13})\,\left[ - 1/12\,R^{p}_{q} R^d_{p} 
\,\chi_{13}^2 + 1/24\,R^d_{p}\,\chi_{13}^2 \right]
\:+\,H^{F'}_{21}(3,\chi_{13},\chi_p,\chi_q;\chi_{13})\,\left[ - 1/12\,R^{q}_{p} R^d_{p} 
\,\chi_{13}^2 \right. \nonumber \\
&+& 1/24 \left. R^d_{p}\,\chi_{13}^2 \right]
\:+\,1/48\,H^{F'}_{21}(7,\chi_{13},\chi_p,\chi_p;\chi_{13})\,(R^{d}_{p})^2\,\chi_{13}^2
\:+\,1/24\,H^{F'}_{21}(7,\chi_{13},\chi_1,\chi_3;\chi_{13})\,R^d_{1} R^d_{3}\,\chi_{13}^2, \\
\label{F0p621loop} 
&& \nonumber \\
\delta^{(6)22}_{\mathrm{loops}} & = & 
  \pi_{16}\,L^r_{0}\,\left[ 4/9\,\chi_\eta \chi_4 - 1/2\,\chi_1 \chi_3 + \chi_{13}^2 
- 13/3\,\bar\chi_{1} \chi_{13} - 35/18\,\bar\chi_{2} \right]
\:-\,2\,\pi_{16}\,L^r_{1}\,\chi_{13}^2 \nonumber \\ 
&-& \pi_{16}\,L^r_{2}\,\left[ 11/3\,\chi_\eta \chi_4 + \chi_{13}^2 
+ 13/3\,\bar\chi_{2} \right] + \pi_{16}\,L^r_{3}\,\left[ 4/9\,\chi_\eta \chi_4 
- 7/12\,\chi_1 \chi_3 + 11/6\,\chi_{13}^2 - 17/6\,\bar\chi_{1} \chi_{13} 
- 43/36\,\bar\chi_{2} \right] \nonumber \\
&+& \pi_{16}^2\,\left[ - 15/64\,\chi_\eta \chi_4 - 59/384\,\chi_1 \chi_3 + 65/384\,\chi_{13}^2 
- 1/2\,\bar\chi_{1} \chi_{13} - 43/128\,\bar\chi_{2} \right]
\:-\,48\,L^r_{4}L^r_{5}\,\bar\chi_{1} \chi_{13}
\:-\,72\,L^{r2}_{4}\,\bar\chi_{1}^2 \nonumber \\
&-& 8\,L^{r2}_{5}\,\chi_{13}^2 
\:+\,\bar{A}(\chi_p)\,\pi_{16}\,\left[ - 1/24\,\chi_p + 1/48\,\bar\chi_{1} 
- 1/8\,\bar\chi_{1}\,R^{p}_{q\eta} + 1/16\,\bar\chi_{1}\,R^c_{p} - 1/48\,R^{p}_{q\eta}\,\chi_p 
- 1/16\,R^{p}_{q\eta}\,\chi_q \right.\nonumber \\ 
&+& \left. 1/48\,R^{\eta}_{pp}\,\chi_\eta + 1/16\,R^c_{p}\,\chi_{13} \right]
\:+\,\bar{A}(\chi_p)\,L^r_{0}\,\left[ 8/3\,R^{p}_{q\eta}\,\chi_p + 2/3\,R^c_{p}\,\chi_p 
+ 2/3\,R^d_{p} \right]  
\:+\,\bar{A}(\chi_p)\,L^r_{3}\,\left[ 2/3\,R^{p}_{q\eta}\,\chi_p 
\right. \nonumber \\
&+& \left. 5/3\,R^c_{p}\,\chi_p + 5/3\,R^d_{p} \right]
\:+\,\bar{A}(\chi_p)\,L^r_{4}\,\left[ - 2\,\bar\chi_{1} \bar\chi^{pp}_{\eta\eta 0} 
- 2\,\bar\chi_{1}\,R^{p}_{q\eta} + 3\,\bar\chi_{1}\,R^c_{p} 
\right]
\:+\, \bar{A}(\chi_p)\,L^r_{5}\,\left[ - 2/3\,\bar\chi^{pp}_{\eta\eta 1} 
- R^{p}_{q\eta}\,\chi_p \right. \nonumber \\
&+& \left. 1/3\,R^{p}_{q\eta}\,\chi_q + 1/2\,R^c_{p}\,\chi_p 
- 1/6\,R^c_{p}\,\chi_q \right]
\:+\, \bar{A}(\chi_p)^2\,\left[ 1/16 + 1/72\,(R^{p}_{q\eta})^2 - 1/72\,R^{p}_{q\eta} R^c_{p} + 
1/288\,(R^{c}_{p})^2 \right] \nonumber \\
&+& \bar{A}(\chi_p) \bar{A}(\chi_{ps})\,\left[ - 1/36\,R^{p}_{q\eta} - 5/72\,R^{p}_{s\eta} 
+ 7/144\,R^c_{p} \right]
\:-\,\bar{A}(\chi_p) \bar{A}(\chi_{qs})\,\left[ 1/36\,R^{p}_{q\eta} + 1/24\,R^{p}_{s\eta} 
+ 1/48\,R^c_{p} \right] \nonumber \\
&+& \bar{A}(\chi_p) \bar{A}(\chi_\eta)\,\left[ - 1/72\,R^{p}_{q\eta} R^v_{\eta 13} 
+ 1/144\,R^c_{p} R^v_{\eta 13} \right]
\:+\,1/8\,\bar{A}(\chi_p) \bar{A}(\chi_{13})
\:+\,1/12\,\bar{A}(\chi_p) \bar{A}(\chi_{46})\,R^{\eta}_{pp} \nonumber \\
&+& \bar{A}(\chi_p) \bar{B}(\chi_p,\chi_p;0)\,\left[ 1/4\,\chi_p
- 1/18\,R^{p}_{q\eta} R^c_{p}\,\chi_p - 1/72\,R^{p}_{q\eta} R^d_{p} + 
1/18\,(R^{c}_{p})^2\,\chi_p + 1/144\,R^c_{p} R^d_{p} \right] \nonumber \\
&+& \bar{A}(\chi_p) \bar{B}(\chi_p,\chi_\eta;0)\,\left[ 1/18\,R^{\eta}_{pp} R^c_{p}\,\chi_p 
- 1/18\,R^{\eta}_{13} R^c_{p}\,\chi_p \right]
\:+\,\bar{A}(\chi_p) \bar{B}(\chi_q,\chi_q;0)\left[ - 1/72\,R^{p}_{q\eta} R^d_{q} 
+ 1/144\,R^c_{p} R^d_{q} \right] \nonumber \\
&-& 1/12\,\bar{A}(\chi_p) \bar{B}(\chi_{ps},\chi_{ps};0)\,R^{p}_{s\eta}\,\chi_{ps}
\:-\,1/18\,\bar{A}(\chi_p) \bar{B}(\chi_1,\chi_3;0)\,R^{q}_{p\eta} R^c_{p}\,\chi_p \nonumber \\
&+& 1/18\,\bar{A}(\chi_p) \bar{C}(\chi_p,\chi_p,\chi_p;0)\,R^c_{p} R^d_{p}\,\chi_p 
\:+\,\bar{A}(\chi_p;\varepsilon)\,\pi_{16}\,\left[ 1/8\,\bar\chi_{1} R^{p}_{q\eta} 
- 1/16 \bar\chi_{1}\,R^c_{p} - 1/16\,R^c_{p}\,\chi_p - 1/16\,R^d_{p} \right] \nonumber \\
&+& \bar{A}(\chi_{ps})\,\pi_{16}\,\left[ 1/16\,\chi_{ps} - 3/16\,\chi_{qs} 
- 3/16\,\bar\chi_{1} \right]
\:-\,2\,\bar{A}(\chi_{ps})\,L^r_{0}\,\chi_{ps}
\:-\,5\,\bar{A}(\chi_{ps})\,L^r_{3}\,\chi_{ps}
\:-\,3\,\bar{A}(\chi_{ps})\,L^r_{4}\,\bar\chi_{1} \nonumber \\
&+& \bar{A}(\chi_{ps})\,L^r_{5}\,\chi_{13}
\:+\,\bar{A}(\chi_{ps}) \bar{A}(\chi_\eta)\,\left[ 7/144\,R^{\eta}_{pp} - 5/72\,R^{\eta}_{ps} 
- 1/48\,R^{\eta}_{qq} + 5/72\,R^{\eta}_{qs} - 1/36\,R^{\eta}_{13} \right] \nonumber \\
&+& \bar{A}(\chi_{ps}) \bar{B}(\chi_p,\chi_p;0)\,\left[1/24\,R^{p}_{s\eta}\,\chi_p 
- 5/24\,R^{p}_{s\eta}\,\chi_{ps} \right]
\:+\,\bar{A}(\chi_{ps}) \bar{B}(\chi_p,\chi_\eta;0)\,\left[ 
- 1/18\,R^{\eta}_{ps} R^z_{qp\eta}\,\chi_p \right. \nonumber \\
&-& 1/9 \left. R^{\eta}_{ps} R^z_{qp\eta}\,\chi_{ps} \right] 
\:-\,1/48\,\bar{A}(\chi_{ps}) \bar{B}(\chi_q,\chi_q;0)\,R^d_{q}
\:+\,1/18\,\bar{A}(\chi_{ps}) \bar{B}(\chi_1,\chi_3;0)\,R^{q}_{s\eta}\,\chi_s \nonumber \\
&+& 1/9\,\bar{A}(\chi_{ps}) \bar{B}(\chi_1,\chi_3;0,k)\,R^{q}_{s\eta}
\:+\,3/16\,\bar{A}(\chi_{ps};\varepsilon)\,\pi_{16}\,\left[\chi_s + \bar\chi_{1} \right] 
\:-\,1/8\,\bar{A}(\chi_{p4})^2 
\:-\,1/8\,\bar{A}(\chi_{p4}) \bar{A}(\chi_{p6}) \nonumber \\
&+& 1/8\,\bar{A}(\chi_{p4}) \bar{A}(\chi_{q6}) 
\:-\,1/32\,\bar{A}(\chi_{p6})^2 
\:+\,\bar{A}(\chi_\eta)\,\pi_{16}\,\left[ 1/16\,\bar\chi_{1}\,R^v_{\eta 13}
- 1/48\,R^v_{\eta 13}\,\chi_\eta + 1/16\,R^v_{\eta 13}\,\chi_{13} \right] \nonumber \\
&+& \bar{A}(\chi_\eta)\,L^r_{0}\,\left[ 4 R^{\eta}_{13}\,\chi_\eta 
+ 2/3\,R^v_{\eta 13}\,\chi_\eta \right] \nonumber
\:-\,8\,\bar{A}(\chi_\eta)\,L^r_{1}\,\chi_\eta
\:-\,2\,\bar{A}(\chi_\eta)\,L^r_{2}\,\chi_\eta
\:+\,\bar{A}(\chi_\eta)\,L^r_{3}\,\left[ 4 R^{\eta}_{13}\,\chi_\eta 
+ 5/3\,R^v_{\eta 13} \chi_\eta \right] \nonumber \\
&+& \bar{A}(\chi_\eta)\,L^r_{4}\,\left[ 4\,\chi_\eta + \bar\chi_{1}\,R^v_{\eta 13} \right] 
\:-\,\bar{A}(\chi_\eta)\,L^r_{5}\,\left[ 1/6\,R^{\eta}_{pp}\,\chi_q 
+ R^{\eta}_{13}\,\chi_{13} + 1/6\,R^v_{\eta 13}\,\chi_\eta \right]
\:+\,1/288\,\bar{A}(\chi_\eta)^2\,(R^v_{\eta 13})^2 \nonumber \\
&+& 1/12\,\bar{A}(\chi_\eta) \bar{A}(\chi_{46})\,R^v_{\eta 13} 
\:+\,\bar{A}(\chi_\eta) \bar{B}(\chi_p,\chi_p;0)\,\left[ - 1/36\,\bar\chi^{pp\eta}_{\eta\eta 1}
- 1/18\,R^{p}_{q\eta} R^{\eta}_{pp}\,\chi_p + 1/18\,R^{\eta}_{pp} R^c_{p}\,\chi_p 
\right. \nonumber \\
&+& 1/144 \left. R^d_{p} R^v_{\eta 13} \right]
\:+\,\bar{A}(\chi_\eta) \bar{B}(\chi_p,\chi_\eta;0)\,\left[ 
- 1/18\,\bar\chi^{\eta p\eta}_{p\eta 1} + 1/18\,\bar\chi^{\eta p\eta}_{q\eta 1}
+ 1/18\,(R^{\eta}_{pp})^2 R^z_{qp\eta}\,\chi_p \right] \nonumber \\ 
&-& 1/12\,\bar{A}(\chi_\eta) \bar{B}(\chi_{ps},\chi_{ps};0)\,R^{\eta}_{ps}\,\chi_{ps}
\:-\,\bar{A}(\chi_\eta) \bar{B}(\chi_\eta,\chi_\eta;0)\,\left[ 1/216\,R^v_{\eta 13}\,\chi_4 
+ 1/27\,R^v_{\eta 13}\,\chi_6 \right] \nonumber \\ 
&-& 1/18\,\bar{A}(\chi_\eta) \bar{B}(\chi_1,\chi_3;0)\,R^1_{\eta\eta} R^3_{\eta\eta}\,\chi_\eta
\:+\,1/18\,\bar{A}(\chi_\eta) \bar{C}(\chi_p,\chi_p,\chi_p;0)\,R^{\eta}_{pp} R^d_{p}\,\chi_p 
\:+\,\bar{A}(\chi_\eta;\varepsilon)\,\pi_{16}\,\left[ 1/8\,\chi_\eta \right. \nonumber \\
&-& 1/16\,\bar\chi_{1}\,R^v_{\eta 13} - 1/8\,R^{\eta}_{13}\,\chi_\eta 
- \left. 1/16\,R^v_{\eta 13}\,\chi_\eta \right]
\:+\,\bar{A}(\chi_1) \bar{A}(\chi_3)\left[ - 1/72\,R^{p}_{q\eta} R^c_{q} + 1/36\,R^{1}_{3\eta} 
R^{3}_{1\eta} + 1/144\,R^c_{1} R^c_{3} \right] \nonumber \\ 
&-& 4\,\bar{A}(\chi_{13})\,L^r_{1}\,\chi_{13}
\:-\,10\,\bar{A}(\chi_{13})\,L^r_{2}\,\chi_{13} 
\:+\,1/8\,\bar{A}(\chi_{13})^2 
\:-\,1/2\,\bar{A}(\chi_{13}) \bar{B}(\chi_1,\chi_3;0,k) \nonumber \\
&+& 1/4\,\bar{A}(\chi_{13};\varepsilon)\,\pi_{16}\,\chi_{13} 
\:+\,1/4\,\bar{A}(\chi_{14}) \bar{A}(\chi_{34})
\:+\,1/16\,\bar{A}(\chi_{16}) \bar{A}(\chi_{36}) 
\:-\,24\,\bar{A}(\chi_4)\,L^r_{1}\,\chi_4
\:-\,6\,\bar{A}(\chi_4)\,L^r_{2}\,\chi_4 \nonumber \\
&+& 12\,\bar{A}(\chi_4)\,L^r_{4}\,\chi_4 
\:+\,1/12\,\bar{A}(\chi_4) \bar{B}(\chi_p,\chi_p;0)\,(R^{p}_{4\eta})^2\,\chi_4 
\:+\,1/6\,\bar{A}(\chi_4) \bar{B}(\chi_p,\chi_\eta;0)\,\left[ R^{p}_{4\eta} 
R^{\eta}_{p4}\,\chi_4 - \,R^{p}_{4\eta} R^{\eta}_{q4}\,\chi_4 \right] \nonumber \\
&-& 1/24\,\bar{A}(\chi_4) \bar{B}(\chi_\eta,\chi_\eta;0)\,R^v_{\eta 13}\,\chi_4
\:-\,1/6\,\bar{A}(\chi_4) \bar{B}(\chi_1,\chi_3;0)\,R^{1}_{4\eta} R^{3}_{4\eta}\,\chi_4
\:+\,3/8\,\bar{A}(\chi_4;\varepsilon)\,\pi_{16}\,\chi_4 \nonumber \\ 
&-& 32\,\bar{A}(\chi_{46})\,L^r_{1}\,\chi_{46}
\:-\,8\,\bar{A}(\chi_{46})\,L^r_{2}\,\chi_{46}
\:+\,16\,\bar{A}(\chi_{46})\,L^r_{4}\,\chi_{46}
\:+\,\bar{A}(\chi_{46}) \bar{B}(\chi_p,\chi_p;0)\,\left[ 1/9\,\chi_{46} + 1/12\,R^{\eta}_{pp} 
\,\chi_p \right. \nonumber \\
&+& 1/36\,R^{\eta}_{pp}\,\chi_4 + \left. 1/9\,R^{\eta}_{p4}\,\chi_6 \right]
\:+\,\bar{A}(\chi_{46}) \bar{B}(\chi_p,\chi_\eta;0)\,\left[ - 1/18\,R^{\eta}_{pp}\,\chi_4 
- 1/9\,R^{\eta}_{p4}\,\chi_6 + 1/9\,R^{\eta}_{q4}\,\chi_6 + 1/18\,R^{\eta}_{13}\,\chi_4 
\right] \nonumber \\
&-& 1/6\,\bar{A}(\chi_{46}) \bar{B}(\chi_p,\chi_\eta;0,k)\,\left[ R^{\eta}_{pp} - R^{\eta}_{13} 
\right] 
\:+\,1/9\,\bar{A}(\chi_{46}) \bar{B}(\chi_\eta,\chi_\eta;0)\,R^v_{\eta 13}\,\chi_{46}
\:-\,\bar{A}(\chi_{46}) \bar{B}(\chi_1,\chi_3;0)\,\left[ 2/9\,\chi_{46} \right. \nonumber \\
&+& 1/9\,R^{\eta}_{p4}\,\chi_6 + \left. 1/18\,R^{\eta}_{13}\,\chi_4 \right] 
\:-\,1/6\,\bar{A}(\chi_{46}) \bar{B}(\chi_1,\chi_3;0,k)\,R^{\eta}_{13}
\:+\,1/2\,\bar{A}(\chi_{46};\varepsilon)\,\pi_{16}\,\chi_{46} \nonumber \\
&+& \bar{B}(\chi_p,\chi_p;0)\,\pi_{16}\,\left[ 1/16\,\bar\chi_{1}\,R^d_{p} + 1/96\,R^d_{p} 
\,\chi_p + 1/32\,R^d_{p}\,\chi_q \right]
\:+\,2/3\,\bar{B}(\chi_p,\chi_p;0)\,L^r_{0}\,R^d_{p}\,\chi_p \nonumber \\
&+& 5/3\,\bar{B}(\chi_p,\chi_p;0)\,L^r_{3}\,R^d_{p}\,\chi_p
\:+\,\bar{B}(\chi_p,\chi_p;0)\,L^r_{4}\,\left[ - 2\,\bar\chi_{1} \bar\chi^{pp}_{\eta\eta 0}
\chi_p - 4\,\bar\chi_{1} R^{p}_{q\eta}\,\chi_p + 4\,\bar\chi_{1} R^c_{p}\,\chi_p 
+ 3\,\bar\chi_{1} R^d_{p} \right] \nonumber \\ 
&+& \bar{B}(\chi_p,\chi_p;0)\,L^r_{5}\,\left[ - 2/3 \bar\chi^{pp}_{\eta\eta 1}\chi_p - 
4/3\,R^{p}_{q\eta}\,\chi_p^2 + 4/3\,R^c_{p}\,\chi_p^2 + 1/2\,R^d_{p}\,\chi_p 
- 1/6\,R^d_{p}\,\chi_q \right] \nonumber \\
&+& \bar{B}(\chi_p,\chi_p;0)\,L^r_{6}\,\left[ 4\,\bar\chi_{1} \bar\chi^{pp}_{\eta\eta 1}
+ 8\,\bar\chi_{1} R^{p}_{q\eta}\,\chi_p - 8\,\bar\chi_{1} R^c_{p}\,\chi_p \right]
\:+\,4\,\bar{B}(\chi_p,\chi_p;0)\,L^r_{7}\,(R^{d}_{p})^2 \nonumber \\
&+& \bar{B}(\chi_p,\chi_p;0)\,L^r_{8}\,\left[ 4/3\,\bar\chi^{pp}_{\eta\eta 2}
+ 8/3\,R^{p}_{q\eta}\,\chi_p^2 - 8/3\,R^c_{p}\,\chi_p^2 \right]
\:+\,\bar{B}(\chi_p,\chi_p;0)^2\,\left[ - 1/18\,R^{p}_{q\eta} R^d_{p}\,\chi_p + 1/18\,R^c_{p} 
R^d_{p}\,\chi_p \right. \nonumber \\
&+& 1/288 \left. (R^{d}_{p})^2 \right]
\:+\,1/18\,\bar{B}(\chi_p,\chi_p;0) \bar{B}(\chi_p,\chi_\eta;0)\,\left[ R^{\eta}_{pp} 
R^d_{p}\,\chi_p - R^{\eta}_{13} R^d_{p}\,\chi_p \right] \nonumber \\
&-& 1/18\,\bar{B}(\chi_p,\chi_p;0) \bar{B}(\chi_1,\chi_3;0)\,R^{q}_{p\eta} R^d_{p}\,\chi_p 
\:+\,1/18\,\bar{B}(\chi_p,\chi_p;0) \bar{C}(\chi_p,\chi_p,\chi_p;0)\,(R^{d}_{p})^2\,\chi_p 
\nonumber \\
&-& 1/16\,\bar{B}(\chi_p,\chi_p;0,\varepsilon)\,\pi_{16}\,\left[ \bar\chi_{1} R^d_{p} 
+ R^d_{p}\,\chi_p \right]
\:+\,8\,\bar{B}(\chi_p,\chi_\eta;0)\,L^r_{7}\,R^z_{qp\eta} R^d_{p} R^z_{\eta 46p}
\nonumber \\
&+& 8/3\,\bar{B}(\chi_p,\chi_\eta;0)\,L^r_{8}\,R^z_{qp\eta} R^d_{p} R^z_{\eta 46p}
\:-\,6\,\bar{B}(\chi_{ps},\chi_{ps};0)\,L^r_{4}\,\bar\chi_{1} \chi_{ps}
\:-\,2\,\bar{B}(\chi_{ps},\chi_{ps};0)\,L^r_{5}\,\chi_{ps}^2 \nonumber \\
&+& 12\,\bar{B}(\chi_{ps},\chi_{ps};0)\,L^r_{6}\,\bar\chi_{1} \chi_{ps}
\:+\,4\,\bar{B}(\chi_{ps},\chi_{ps};0)\,L^r_{8}\,\chi_{ps}^2
\:+\,2\,\bar{B}(\chi_\eta,\chi_\eta;0)\,L^r_{4}\,\bar\chi_{1} R^v_{\eta 13}\,\chi_\eta
\nonumber \\ 
&+& 2/3\,\bar{B}(\chi_\eta,\chi_\eta;0)\,L^r_{5}\,R^v_{\eta 13}\,\chi_\eta^2 
\:-\,4\,\bar{B}(\chi_\eta,\chi_\eta;0)\,L^r_{6}\,\bar\chi_{1} R^v_{\eta 13}\,\chi_\eta
\:+\,4\,\bar{B}(\chi_\eta,\chi_\eta;0)\,L^r_{7}\,R^z_{311\eta\eta}(R^z_{\eta 461})^2 
\nonumber \\
&-& \bar{B}(\chi_\eta,\chi_\eta;0)\,L^r_{8}\,\left[ 4/9\,R^v_{\eta 13}\,\chi_4^2 
+ 8/9\,R^v_{\eta 13}\,\chi_6^2 \right]
\:+\,1/144\,\bar{B}(\chi_1,\chi_1;0) \bar{B}(\chi_3,\chi_3;0)\,R^d_{1} R^d_{3} \nonumber \\
&-& 8\,\bar{B}(\chi_1,\chi_3;0)\,L^r_{7}\,R^d_{1} R^d_{3}
\:-\,8/3\,\bar{B}(\chi_1,\chi_3;0)\,L^r_{8}\,R^d_{1} R^d_{3}
\:+\,4\,\bar{C}(\chi_p,\chi_p,\chi_p;0)\,L^r_{4}\,\bar\chi_{1} R^d_{p}\,\chi_p 
\nonumber \\
&+& 4/3\,\bar{C}(\chi_p,\chi_p,\chi_p;0)\,L^r_{5}\,R^d_{p}\,\chi_p^2
\:-\,8\,\bar{C}(\chi_p,\chi_p,\chi_p;0)\,L^r_{6}\,\bar\chi_{1} R^d_{p}\,\chi_p 
\:-\,8/3\,\bar{C}(\chi_p,\chi_p,\chi_p;0)\,L^r_{8}\,R^d_{p}\,\chi_p^2 \nonumber \\
&+& H^{F}(1,\chi_p,\chi_p,\chi_{13};\chi_{13})\,\left[ 1/8\,\chi_p - 1/16\,\chi_{13} 
- 1/36\,(R^{p}_{q\eta})^2\,\chi_{13} + 1/36\,R^{p}_{q\eta} R^c_{p}\,\chi_{13} 
- 1/144\,(R^{c}_{p})^2\,\chi_{13} \right] \nonumber \\
&+& H^{F}(1,\chi_p,\chi_{1s},\chi_{3s};\chi_{13})\,\left[ - 1/12\,R^{p}_{q\eta}\,\chi_{qs} 
+ 1/24\,R^{p}_{q\eta}\,\chi_{13} - 1/16\,R^{p}_{s\eta}\,\chi_p + 1/48\,R^{p}_{s\eta}\,\chi_q 
+ 1/24\,R^c_{p}\,\chi_{ps} \right] \nonumber \\
&-& 1/4\,H^{F}(1,\chi_{p4},\chi_{q6},\chi_{46};\chi_{13})\,\chi_{46}
\:+\,H^{F}(1,\chi_\eta,\chi_p,\chi_{13};\chi_{13})\,\left[ 1/36\,R^{p}_{q\eta} R^v_{\eta 13}
\,\chi_{13} - 1/72\,R^c_{p} R^v_{\eta 13}\,\chi_{13} \right] \nonumber \\
&-& 1/144\,H^{F}(1,\chi_\eta,\chi_\eta,\chi_{13};\chi_{13})\,(R^v_{\eta 13})^2\,\chi_{13} 
\:-\,1/48\,H^{F}(1,\chi_\eta,\chi_{1s},\chi_{3s};\chi_{13})\,\left[ R^{\eta}_{ps}\,\chi_p 
- R^{\eta}_{ps}\,\chi_q \right. \nonumber \\
&-& R^v_{\eta ps}\,\chi_\eta - R^v_{\eta 13} \left. \chi_s \right]
\:+\,H^{F}(1,\chi_1,\chi_{13},\chi_3;\chi_{13})\,\left[ 1/36\,R^{p}_{q\eta} R^c_{q}\,\chi_{13} 
- 1/18\,R^{1}_{3\eta} R^{3}_{1\eta}\,\chi_{13} \right. \nonumber \\
&-& 1/72 \left. R^c_{1} R^c_{3}\,\chi_{13} \right]
\:-\,1/8\,H^{F}(1,\chi_{13},\chi_{13},\chi_{13};\chi_{13})\,\chi_{13}
\:-\,3/8\,H^{F}(1,\chi_{14},\chi_{34},\chi_4;\chi_{13})\,\chi_4 \nonumber \\ 
&+& H^{F}(2,\chi_p,\chi_p,\chi_{13};\chi_{13})\,\left[ 1/36\,R^{p}_{q\eta} R^d_{p}\,\chi_{13} 
- 1/72\,R^c_{p} R^d_{p}\,\chi_{13} \right]
\:-\,1/72\,H^{F}(2,\chi_p,\chi_\eta,\chi_{13};\chi_{13})\,R^d_{p} R^v_{\eta 13}\,\chi_{13} 
\nonumber \\
&+& H^{F}(2,\chi_p,\chi_{13},\chi_q;\chi_{13})\,\left[ 1/36\,R^{q}_{p\eta} R^d_{p}\,\chi_{13} 
- 1/72\,R^c_{q} R^d_{p}\,\chi_{13} \right]
\:+\,1/24\,H^{F}(2,\chi_p,\chi_{1s},\chi_{3s};\chi_{13})\,R^d_{p}\,\chi_{ps} \nonumber \\ 
&-& 1/144\,H^{F}(5,\chi_p,\chi_p,\chi_{13};\chi_{13})\,(R^{d}_{p})^2\,\chi_{13}
\:-\,1/72\,H^{F}(5,\chi_1,\chi_3,\chi_{13};\chi_{13})\,R^d_{1} R^d_{3}\,\chi_{13} \nonumber \\
&+& H^{F'}(1,\chi_p,\chi_{1s},\chi_{3s};\chi_{13})\,\left[ - 5/24\,R^{p}_{q\eta}\,\chi_q 
\chi_{13} + 1/12\,R^{p}_{q\eta}\,\chi_{ps} \chi_{13} + 1/16\,R^{p}_{s\eta}\,\chi_q \chi_{13} 
+ 7/48\,R^{p}_{s\eta}\,\chi_{13} \chi_s \right. \nonumber \\
&-& 1/24 \left. R^c_{p}\,\chi_{ps} \chi_{13} \right]
\:+\,1/4\,H^{F'}(1,\chi_{p4},\chi_{q6},\chi_{46};\chi_{13})\,\chi_{13} \chi_{46}
\:+\,H^{F'}(1,\chi_\eta,\chi_p,\chi_{13};\chi_{13})\left[ 1/9\,R^{p}_{q\eta} R^{\eta}_{13} 
\,\chi_{13}^2 \right. \nonumber \\
&-& 1/36 \left. R^{p}_{q\eta} R^v_{\eta 13}\,\chi_{13}^2 
+ 1/18\,R^{\eta}_{pp} R^c_{p}\,\chi_{13}^2 
+ 1/72\,R^c_{p} R^v_{\eta 13}\,\chi_{13}^2 \right]
\:+\,H^{F'}(1,\chi_\eta,\chi_\eta,\chi_{13};\chi_{13})\,\left[ 1/9\,(R^{\eta}_{13})^2 
\,\chi_{13}^2 \right. \nonumber \\ 
&+& 1/9 \left. R^{\eta}_{13} R^v_{\eta 13}\,\chi_{13}^2 + 5/144\,(R^v_{\eta 13})^2\, 
\chi_{13}^2 \right]
\:+\,H^{F'}(1,\chi_\eta,\chi_{1s},\chi_{3s};\chi_{13})\,\left[ - 1/16\,R^{\eta}_{ps}\, 
\chi_p \chi_{13} - 5/48\,R^{\eta}_{ps}\,\chi_q \chi_{13} \right. \nonumber \\
&+& 1/6 \left. R^{\eta}_{1s} R^z_{\eta s3}\,\chi_{13}^2 
- 1/48\,R^v_{\eta ps}\,\chi_\eta \chi_{13} - 1/48\,R^v_{\eta 13}\,\chi_{13} \chi_s \right]
\:+\,H^{F'}(1,\chi_1,\chi_{13},\chi_3;\chi_{13})\,\left[ - 1/36\,R^{p}_{q\eta} R^c_{q} 
\,\chi_{13}^2 \right. \nonumber \\ 
&+& 1/6 \left. R^{1}_{3\eta} R^{3}_{1\eta}\,\chi_{13}^2 + 1/72\,R^c_{1} R^c_{3}\,\chi_{13}^2 
\right]
\:+\,H^{F'}(1,\chi_{13},\chi_p,\chi_p;\chi_{13})\,\left[ - 1/8\,\chi_p \chi_{13} 
- 3/16\,\chi_{13}^2 - 1/36\,(R^{p}_{q\eta})^2\,\chi_{13}^2 \right. \nonumber \\
&+& 1/12 \left. R^{p}_{q\eta} R^c_{p}\,\chi_{13}^2 - 1/48\,(R^{c}_{p})^2\,\chi_{13}^2 \right] 
\:+\,1/8\,H^{F'}(1,\chi_{13},\chi_{13},\chi_{13};\chi_{13})\,\chi_{13}^2 \nonumber \\
&+& 3/8\,H^{F'}(1,\chi_{14},\chi_{34},\chi_4;\chi_{13})\,\chi_{13} \chi_4
\:+\,H^{F'}(2,\chi_p,\chi_p,\chi_{13};\chi_{13})\,\left[ - 1/36\,R^{p}_{q\eta} R^d_{p}\, 
\chi_{13}^2 + 5/72\,R^c_{p} R^d_{p}\,\chi_{13}^2 \right] \nonumber \\
&+& H^{F'}(2,\chi_p,\chi_\eta,\chi_{13};\chi_{13})\,\left[ 1/18\,R^{\eta}_{pp} R^d_{p}\, 
\chi_{13}^2 + 1/72\,R^d_{p} R^v_{\eta 13}\,\chi_{13}^2 \right] \nonumber \\
&+& H^{F'}(2,\chi_p,\chi_{13},\chi_q;\chi_{13})\,\left[ - 1/36\,R^{q}_{p\eta} R^d_{p}\, 
\chi_{13}^2 + 1/72\,R^c_{q} R^d_{p}\,\chi_{13}^2 \right]
\:-\,1/24\,H^{F'}(2,\chi_p,\chi_{1s},\chi_{3s};\chi_{13})\,R^d_{p}\,\chi_{ps} \chi_{13}
\nonumber \\
&+& 5/144\,H^{F'}(5,\chi_p,\chi_p,\chi_{13};\chi_{13})\,(R^{d}_{p})^2\,\chi_{13}^2 
\:+\,1/72\,H^{F'}(5,\chi_1,\chi_3,\chi_{13};\chi_{13})\,R^d_{1} R^d_{3}\,\chi_{13}^2
\nonumber \\ 
&+& H^{F'}_1(1,\chi_p,\chi_p,\chi_{13};\chi_{13})\,\left[ \chi_{13}^2 + 2/9\,(R^{p}_{q\eta})^2 
\,\chi_{13}^2 - 2/9\,R^{p}_{q\eta} R^c_{p}\,\chi_{13}^2 + 1/9\,(R^{c}_{p})^2\,\chi_{13}^2 
\right] \nonumber \\
&+& 1/3\,H^{F'}_1(1,\chi_p,\chi_{1s},\chi_{3s};\chi_{13})\,R^{p}_{q\eta} 
R^z_{sqp}\,\chi_{13}^2
\:-\,1/3\,H^{F'}_1(1,\chi_{ps},\chi_{qs},\chi_p;\chi_{13})\,R^{p}_{s\eta} 
\,\chi_{13}^2 \nonumber \\ 
&-& 1/3\,H^{F'}_1(1,\chi_{ps},\chi_{qs},\chi_\eta;\chi_{13})\,R^{\eta}_{13} 
R^z_{sp\eta}\,\chi_{13}^2
\:+\,1/9\,H^{F'}_1(1,\chi_{13},\chi_p,\chi_\eta;\chi_{13})\,\left[ R^{p}_{q\eta}
R^v_{\eta 13}\,\chi_{13}^2 \right. \nonumber \\
&-& R^{\eta}_{pp} R^z_{qp\eta} R^c_{p} \left. \chi_{13}^2 \right]
\:-\,H^{F'}_1(1,\chi_{13},\chi_\eta,\chi_\eta;\chi_{13})\,\left[ 1/9\,R^{\eta}_{13}
R^v_{\eta 13}\,\chi_{13}^2 + 1/18\,(R^v_{\eta 13})^2\,\chi_{13}^2 \right] \nonumber \\
&+& H^{F'}_1(1,\chi_{13},\chi_1,\chi_3;\chi_{13})\,\left[ 1/9\,R^{p}_{q\eta} R^c_{q}\, 
\chi_{13}^2 - 2/9\,R^{1}_{3\eta} R^{3}_{1\eta}\,\chi_{13}^2 \right]
\:+\,1/9\,H^{F'}_1(3,\chi_{13},\chi_p,\chi_p;\chi_{13})\,\left[ R^{p}_{q\eta} R^d_{p} 
\,\chi_{13}^2 \right. \nonumber \\
&-& R^c_{p} R^d_{p} \left. \chi_{13}^2 \right]
\:+\,1/9\,H^{F'}_1(3,\chi_{13},\chi_p,\chi_q;\chi_{13})\,R^{q}_{p\eta} R^d_{p}\,\chi_{13}^2 
\:+\,1/9\,H^{F'}_1(3,\chi_{13},\chi_p,\chi_\eta;\chi_{13})\,R^{\eta}_{13} R^z_{pq\eta} 
R^d_{p}\,\chi_{13}^2 \nonumber \\
&-& 1/18\,H^{F'}_1(7,\chi_{13},\chi_p,\chi_p;\chi_{13})\,(R^{d}_{p})^2\,\chi_{13}^2 
\:-\,3/8\,H^{F'}_{21}(1,\chi_p,\chi_p,\chi_{13};\chi_{13})\,\chi_{13}^2 \nonumber \\
&-& 1/8\,H^{F'}_{21}(1,\chi_p,\chi_{1s},\chi_{3s};\chi_{13})\,R^{p}_{q\eta} R^z_{sqp}
\,\chi_{13}^2 
\:+\,1/8\,H^{F'}_{21}(1,\chi_{ps},\chi_{qs},\chi_p;\chi_{13})\,\left[ R^{p}_{q\eta}\, 
\chi_{13}^2 + R^{p}_{s\eta}\,\chi_{13}^2 - R^c_{p}\,\chi_{13}^2 \right] \nonumber \\
&+& 1/8\,H^{F'}_{21}(1,\chi_{ps},\chi_{qs},\chi_q;\chi_{13})\,R^{q}_{p\eta} 
R^z_{spq}\,\chi_{13}^2
\:+\,1/8\,H^{F'}_{21}(1,\chi_{ps},\chi_{qs},\chi_\eta;\chi_{13})\,R^{\eta}_{13} 
R^z_{pq\eta} R^z_{sp\eta}\,\chi_{13}^2 \nonumber \\
&-& 1/8\,H^{F'}_{21}(1,\chi_\eta,\chi_{1s},\chi_{3s};\chi_{13})\,R^{\eta}_{13} 
R^z_{s1\eta} R^z_{s3\eta}\,\chi_{13}^2
\:+\,H^{F'}_{21}(1,\chi_{13},\chi_p,\chi_p;\chi_{13})\,\left[  3/16\,\chi_{13}^2 
+ 1/12\,(R^{p}_{q\eta})^2\,\chi_{13}^2 \right. \nonumber \\
&-& 1/12 \left. R^{p}_{q\eta} R^c_{p}\,\chi_{13}^2 
+ 1/48\,(R^{c}_{p})^2\,\chi_{13}^2 \right]
\:+\,H^{F'}_{21}(1,\chi_{13},\chi_p,\chi_\eta;\chi_{13})\,\left[ - 1/12\,R^{p}_{q\eta} 
R^v_{\eta 13}\,\chi_{13}^2 + 1/24\,R^c_{p} R^v_{\eta 13}\,\chi_{13}^2 \right] \nonumber \\
&+& 1/48\,H^{F'}_{21}(1,\chi_{13},\chi_\eta,\chi_\eta;\chi_{13})\,(R^v_{\eta 
13})^2\,\chi_{13}^2 
\:+\,H^{F'}_{21}(1,\chi_{13},\chi_1,\chi_3;\chi_{13})\,\left[ - 1/12\,R^{p}_{q\eta} R^c_{q} 
\,\chi_{13}^2 + 1/6\,R^{1}_{3\eta} R^{3}_{1\eta}\,\chi_{13}^2 \right. \nonumber \\
&+& 1/24 \left. R^c_{1} R^c_{3} \,\chi_{13}^2 \right]
\:+\,3/8\,H^{F'}_{21}(1,\chi_{13},\chi_{13},\chi_{13};\chi_{13})\,\chi_{13}^2
\:+\,9/8\,H^{F'}_{21}(1,\chi_4,\chi_{14},\chi_{34};\chi_{13})\,\chi_{13}^2 \nonumber \\
&+& 3/4\,H^{F'}_{21}(1,\chi_{46},\chi_{p4},\chi_{q6};\chi_{13})\,\chi_{13}^2 
\:-\,1/8\,H^{F'}_{21}(3,\chi_{ps},\chi_p,\chi_{qs};\chi_{13})\,R^d_{p}\,\chi_{13}^2
\nonumber \\
&+& H^{F'}_{21}(3,\chi_{13},\chi_p,\chi_p;\chi_{13})\,\left[ - 1/12\,R^{p}_{q\eta} R^d_{p} 
\,\chi_{13}^2 + 1/24\,R^c_{p} R^d_{p}\,\chi_{13}^2 \right] 
\:+\,H^{F'}_{21}(3,\chi_{13},\chi_p,\chi_q;\chi_{13})\,\left[ - 1/12\,R^{q}_{p\eta} R^d_{p} 
\,\chi_{13}^2 \right. \nonumber \\
&+& 1/24 \left. R^c_{q} R^d_{p}\,\chi_{13}^2 \right]
\:+\,1/24\,H^{F'}_{21}(3,\chi_{13},\chi_p,\chi_\eta;\chi_{13})\,R^d_{p} R^v_{\eta 13}
\,\chi_{13}^2
\:+\,1/48\,H^{F'}_{21}(7,\chi_{13},\chi_p,\chi_p;\chi_{13})\,(R^{d}_{p})^2\,\chi_{13}^2 
\nonumber \\
&+& 1/24\,H^{F'}_{21}(7,\chi_{13},\chi_1,\chi_3;\chi_{13})\,R^d_{1} R^d_{3}\,\chi_{13}^2.
\label{F0p622loop} 
\end{eqnarray} 

\end{widetext}

\section{Discussion and Conclusions}

The analytical formulas for the NNLO contributions to the decay
constants are very complicated, especially when the valence quark masses 
are nondegenerate. The practical usefulness of those results thus depends 
completely on the availability of convenient numerical implementations.
In the near future, we will make such an implementation
available~\cite{website}.

The results, as given in the preceding section, depend on nine
${\cal O}(p^4)$ and five ${\cal O}(p^6)$ LEC:s.
Eventually, their proper values should be
determined from lattice QCD by a fit of the NLO + NNLO formulas to the
simulation data. At the present time, these input parameters have been 
taken from the continuum work in NNLO $\chi$PT of Ref.~\cite{ABT2}. 
The combination of LEC:s used in this paper corresponds to 'fit~10' of
Ref.~\cite{ABT2}, which had $F_0 = 87.7$ MeV and $\mu = 770$ MeV. 
The parameters which were 
not determined by that fit, namely $L^r_0,L^r_4$ and $L^r_6$, 
have been set to zero, as have 
all the $K^r_i$. This has been done at a scale of $\mu=~770$~MeV.
It should be noted that $L^r_0$ cannot be determined from experiment, but is
obtainable from partially quenched simulations.
Some recent results on $L^r_4$ and $L^r_6$ may be found in Ref.~\cite{piK}.

The graphical presentation of multidimensional functions is a more immediate
problem. In the most general case, the decay constant of a charged 
pseudoscalar meson in PQ$\chi$PT is a function of two valence and three sea 
quark masses. The plots presented in this section 
should therefore be viewed as an attempt to cover the potentially most 
interesting parts of the parameter space. They also serve as a consistency
check for the formulas in the preceding section, as the cases with fewer 
different masses should be obtained numerically as limits of the more 
general cases.

\subsection{Numerical Results for $d_{\mathrm{sea}} = 1$}

\begin{figure}[h!t]
\begin{center}
\includegraphics[width=7.8cm]{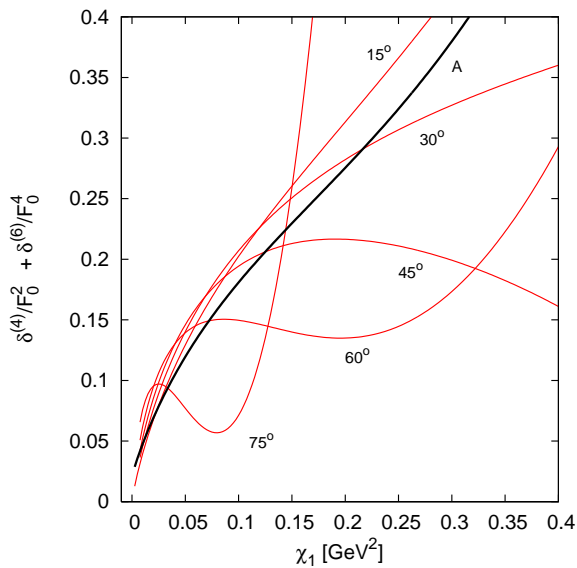}
\caption{The combined NLO and NNLO shifts of the decay constant, plotted for 
$d_{\mathrm{val}} = 1$ and $d_{\mathrm{sea}} = 1$, for values of $\theta$ 
between $15^\circ$ and $75^\circ$. The curve 'A' corresponds to 
$\chi_4 = 0.125\:\mathrm{GeV}^2$. In order to reduce clutter, most of the lines
in this plot, with the exception of the $15^\circ$ line, have not been drawn 
all the way to the origin.}
\label{11fig}
\end{center}
\end{figure}

To this end, the result for $d_{\mathrm{val}} = 1$ 
and $d_{\mathrm{sea}} = 1$ has been plotted along different rays in the
$\chi_1$-$\chi_4$ 
plane, characterized by an angle $\theta$ which is defined according to
\begin{eqnarray}
\tan(\theta) &=& \chi_4 / \chi_1,
\label{rayplot}
\end{eqnarray}
such that $\theta = 45^\circ$ corresponds to the unquenched case 
of equal sea and valence quark 
masses. The typical situation in lattice QCD simulations, where sea
quarks are heavier than their valence counterparts, is attained 
for $\theta > 45^\circ$.
The shift in the decay constant due to the combined NLO + NNLO 
contributions has been plotted in Fig.~\ref{11fig} for 
different values of $\theta$. The quantity 
plotted, $\Delta$, denotes the relative change in the decay constant and
is defined in accordance with Eq.~(\ref{delteq}), as
\begin{eqnarray}
F_{\mathrm{phys}} &=& F_0\,\left(1 + \Delta \right).
\end{eqnarray}
Thus, when plotted along a curve of the form~(\ref{rayplot}), $\Delta$
is expected to vanish in the chiral limit, which is indeed borne out 
in all plots presented in this section.
This shows that the NNLO shifts to the decay constant in PQ$\chi$PT
do not produce unphysical logarithms.
On the other hand, the curve labeled 'A' in Fig.~\ref{11fig} is plotted for a 
constant sea mass with $\chi_4 = 0.125\:\mathrm{GeV}^2$, and in that
case $\Delta$ approaches a constant as $\chi_1 \rightarrow 0$. 

\begin{figure}[h!t]
\begin{center}
\includegraphics[width=7.8cm]{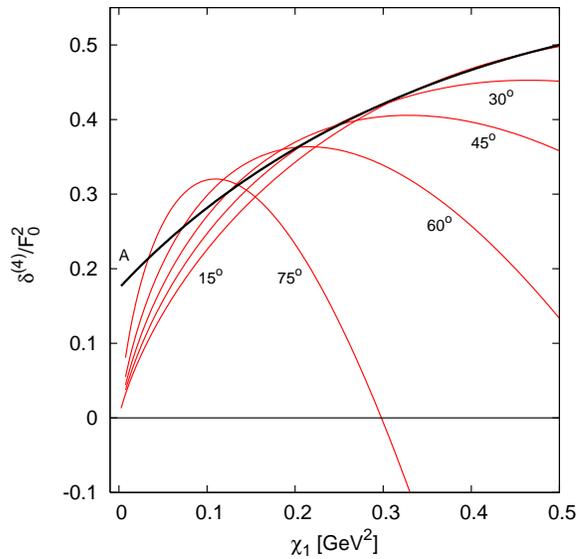}
\caption{The NLO shift of the decay constant, plotted for 
$d_{\mathrm{val}} = 1$ and $d_{\mathrm{sea}} = 1$,
for values of $\theta$ between
$15^\circ$ and $75^\circ$.
The curve 'A' corresponds to $\chi_4 = 0.125\:\mathrm{GeV}^2$.}
\label{11bfig}
\end{center}
\end{figure}

\begin{figure}[h!t]
\begin{center}
\includegraphics[width=7.8cm]{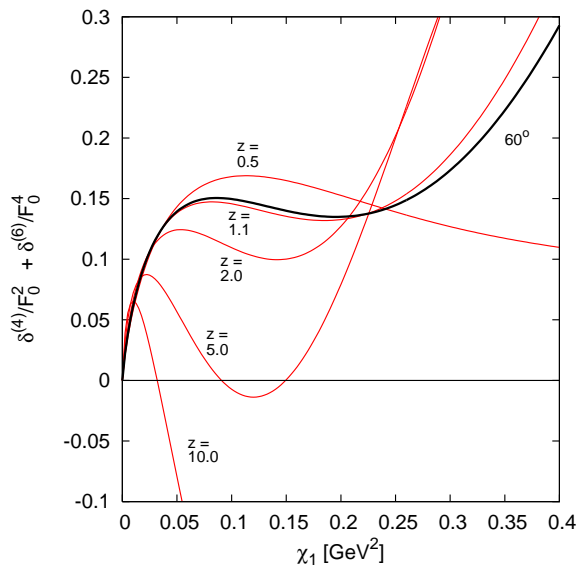}
\caption{The combined NLO and NNLO shifts of the decay constant, plotted for 
$d_{\mathrm{val}} = 1$, $d_{\mathrm{sea}} = 2$, $\theta = 60^\circ$, and a strange sea
quark mass parameter $z$ between $1/2$ and $10$.}
\label{12fig}
\end{center}
\end{figure}

The results along the lines~(\ref{rayplot}) at NLO is shown in Fig.~\ref{11bfig}.
The chiral expansion for the decay constants converges better than for
the masses as can be seen by comparison of Figs.~\ref{11fig} and~\ref{11bfig}
with the
corresponding ones in Ref.~\cite{BDL}. Note, however, that the dependence of
$\Delta$ on the numerical values of the LEC:s is strong and can change the behavior 
of the NNLO results quite considerably. It should also be kept in mind that
many of the curves plotted in Fig.~\ref{11fig} reach far beyond the 
expected radius of convergence of PQ$\chi$PT.

\subsection{Numerical Results for $d_{\mathrm{sea}} = 2$}

The physically more 
interesting case of $d_{\mathrm{sea}} = 2$, where the strange sea quark mass
may deviate from that of the $u,d$ quarks, can then be accounted 
for by the introduction of an additional parameter $z$, which is defined as
\begin{eqnarray}
z &=& \chi_6\,/\chi_4.
\end{eqnarray}

The change in $\Delta$ when the strange sea quark mass is allowed
to vary is shown in Fig.~\ref{12fig}. In general,
the NNLO effects appear to become larger as the strange sea quark mass is 
increased. Fig.~\ref{12fig} also shows that Eq.~(\ref{F0p612loop})
reduces numerically to the 
$d_{\mathrm{val}} = 1$, $d_{\mathrm{sea}} = 1$ result
of Eq.~(\ref{F0p611loop}) when all sea 
quark masses become equal, i.e. $z \rightarrow 1$.

\begin{figure}[h!t]
\begin{center}
\includegraphics[width=7.8cm]{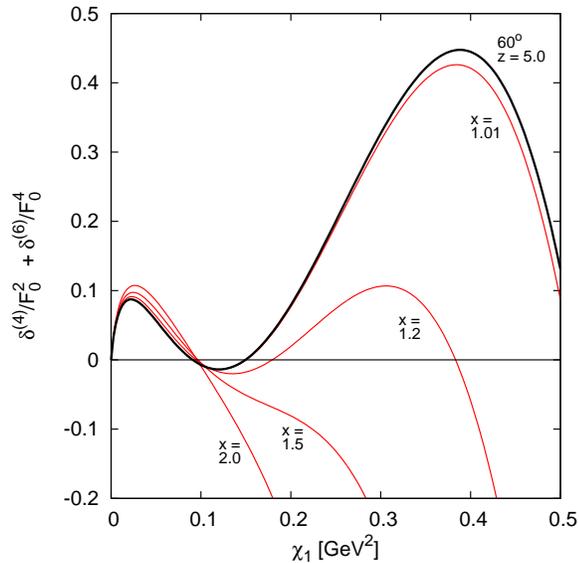}
\caption{The combined NLO and NNLO shifts of the decay constant, plotted for 
$d_{\mathrm{val}} = 2$, $d_{\mathrm{sea}} = 2$, $z = 5$ and $\theta = 60^\circ$. 
The strange valence quark mass parameter $x$ varies between $1$ and $2$.}
\label{22fig}
\end{center}
\end{figure}

It is also instructive to investigate the NNLO correction to the
decay constant for $d_{\mathrm{val}} = 2$ and $d_{\mathrm{sea}} = 2$. 
The behavior of $\Delta$ when the strange valence
quark mass deviates from that of the $u,d$ quarks is shown in Fig.~\ref{22fig},
where the results have been parameterized in terms of
\begin{eqnarray}
x &=& \chi_3\,/\chi_1.
\end{eqnarray}
It is evident that $\Delta$ changes rapidly with increasing $x$ for values of
$\chi_1$ above about $0.15\:\mathrm{GeV}^2$. Also, Eq.~(\ref{F0p622loop}) is 
seen to reduce correctly to the $d_{\mathrm{val}} = 1$, $d_{\mathrm{sea}} = 2$
result when $x \rightarrow 1$.

\subsection{Numerical Results for $d_{\mathrm{sea}} = 3$}

\begin{figure}[h!t]
\begin{center}
\includegraphics[width=7.8cm]{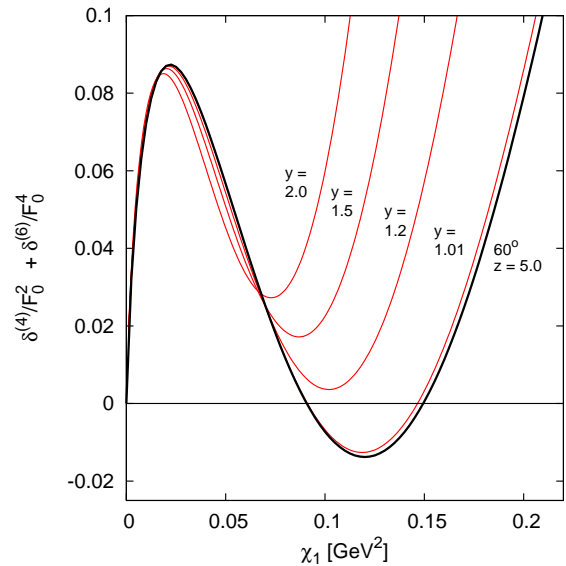}
\caption{The combined NLO and NNLO shifts of the decay constant, plotted for 
$d_{\mathrm{val}} = 1$, $d_{\mathrm{sea}} = 3$, $\theta = 60^\circ$ and $z = 5$. 
The mass parameter $y$ of the down quark in the sea sector ranges between
$1$ and $2$.}
\label{13fig}
\end{center}
\end{figure}

In the case of $d_{\mathrm{sea}} = 3$, $\chi_\pi$ and $\chi_\eta$ are 
non-trivially related to the sea quark masses, and consequently the sea quark 
sector is more complicated. The variation in the mass of the $d$ quark
with respect to the $u$ quark, in the sea quark sector, has been parameterized by
\begin{eqnarray}
y &=& \chi_5\,/\chi_4.
\end{eqnarray}
Sample plots of $\Delta$ for $d_{\mathrm{val}} = 1$, with all three sea quark
masses 
different, are shown in Fig.~\ref{13fig}. Those plots also demonstrate
numerically the 
consistency of Eq.~(\ref{F0p613loop}) with
the $d_{\mathrm{sea}} = 1,2$ results. 

\subsection{Conclusions}

A generic feature of the plots presented in this section is that the 
curves for $\Delta$ show a pronounced dip around $\chi_1 = 0.1\:\mathrm{GeV}^2$.
This indicates a strong cancellation between the NLO and NNLO contributions 
to $\Delta$. It should be noted that such a feature is 
apparently not exhibited by the analogous expressions for the
pseudoscalar meson 
mass~\cite{BDL}. This cancellation also depends strongly on the choice
of the LEC:s.
Many more plots can of course be produced but those presented
give a first indication of the size of the corrections and how they vary
with the different quark masses used as input.
We have not attempted any fit of our results to the available lattice data;
Many simulations are performed with only two flavors of sea quarks, and in 
addition we need extrapolations to zero lattice spacing and infinite volume at 
each quark mass to apply the present formulas. Work for the case with two sea
quark flavors is in progress.

In conclusion, we have calculated the decay constants of the charged,
or off-diagonal, 
pseudoscalar mesons to NNLO in PQ$\chi$PT and
presented analytical as well as numerical 
results for a variety of different combinations of quark masses.
The NNLO contributions were 
found, as expected from previous work in NNLO $\chi$PT \cite{ABT1,ABT2,KG},
to be sizable even though 
there is a tendency toward cancellation with the NLO result.
As the results depend on a number 
of largely unknown LEC:s, statements about the convergence of the
chiral expansion have to be 
postponed at this time. 
The NNLO effects are definitely non-negligible at presently used quark 
masses in Lattice QCD simulations.

\section*{Acknowledgments}

The program \verb;FORM 3.0; has been used extensively in these calculations
\cite{FORM}. This work is supported by the European Union TMR network,
Contract No. 
HPRN-CT-2002-00311  (EURIDICE). TL wishes to thank the Thomas
Jefferson National Accelerator 
Facility (TJNAF) and the Helsinki Institute of Physics (HIP),
where part of this paper was 
completed, for their hospitality. TL also thanks the
Mikael Bj\"ornberg memorial foundation 
for a travel grant.

\end{document}